\documentclass[
reprint,
superscriptaddress,
 amsmath,amssymb,
 aps,
floatfix,
]{revtex4-2}

\usepackage{graphicx}
\usepackage{dcolumn}
\usepackage{bm}
\usepackage[colorlinks=true,linkcolor=cyan,urlcolor=cyan,citecolor=cyan]{hyperref}
\usepackage{tikz}
\usetikzlibrary{arrows.meta, positioning} 

\usepackage{comment}

\begin{document}

\preprint{APS/123-QED}

\title{Unravelling the Flow of Information in a Nonequilibrium Process in the Presence of Hydrodynamic Interactions} 

\author{Biswajit Das}
\email{bd18ip005@iiserkol.ac.in}
\affiliation{Department of Physical Sciences, Indian Institute of Science Education and Research Kolkata, Mohanpur Campus, Mohanpur, West Bengal 741246, India}

\author{Sreekanth K Manikandan}
\email{sreekanth.manikandan@physics.gu.se}
\affiliation{Department of Physics, Gothenburg University, Gothenburg, Sweden}

\author{Ayan Banerjee}
\email{ayan@iiserkol.ac.in}
\affiliation{Department of Physical Sciences, Indian Institute of Science Education and Research Kolkata, Mohanpur Campus, Mohanpur, West Bengal 741246, India}

\date{\today}

\begin{abstract}
Identifying the origin of nonequilibrium characteristics in a generic interacting system having multiple degrees of freedom is a challenging task.  In this context, information-theoretic measures such as mutual information and related polymorphs offer valuable insights. Here, we explore these measures in a minimal experimental model consisting of two hydrodynamically coupled colloidal particles, where a nonequilibrium drive is introduced via an exponentially correlated noise acting on one of the particles.  We show that the information-theoretic tools considered enable a systematic, data-driven dissection of information flow within the system. These measures allow us to identify the driving node and reconstruct the directional dependencies between particles. Notably, they help explain a recently observed, counterintuitive trend in the dependence of irreversibility on interaction strength under coarse-graining (\textcolor{cyan}{\href{https://doi.org/10.48550/arXiv.2405.00800}{B. Das et.al., arXiv:2405.00800 (2024)}}). Finally, our results demonstrate how directional information measures can uncover the hidden structure of nonequilibrium dynamics and provide a framework for studying similar effects in more complex systems.

\end{abstract}

\maketitle

%


\section{Introduction}
Understanding how interactions shape dynamics and thermodynamics in complex mesoscopic systems remains a central challenge in nonequilibrium physics~\cite{bo2017multiple, busiello2020coarse,loos2020irreversibility,nicoletti2024informationprx}. When such systems operate far from equilibrium, especially in high-dimensional settings, interactions -- both internal and environmentally mediated -- can lead to emergent behaviors not directly predictable from individual components~\cite{fruchart2021non,cates2015motility}. These emergent features include collective motion~\cite{cates2015motility,bialek2012statistical}, pattern formation~\cite{falasco2018information,brauns2020phase}, and selective amplification of certain pathways or states~\cite{busiello2021dissipation,liang2024thermodynamic}. In realistic settings, the environment introduces additional couplings between subsystems, often with memory or spatial correlations, complicating the interpretation of dynamical dependencies~\cite{hilfinger2011separating,tsimring2014noise,bialek2012statistical,chechkin2017brownian}. This makes it particularly difficult to isolate the origin of nonequilibrium driving or to identify which variables act as effective degrees of freedom in generating or transmitting nonequilibrium features. A fundamental open question, therefore, is: How can one systematically disentangle internal dynamics from environmental influences to identify the true source of irreversibility and driving within complex systems?\\

 Recent developments in information theory have provided promising tools to address this. A prominent measure is mutual information (MI), which quantifies the statistical dependencies between variables~\cite{shannon1948mathematical,cover2005elements}. This has been successfully used to quantify interactions in mesoscopic systems, both at and away from equilibrium~\cite{nicoletti2023informationcriticalitycomplexstochastic}. Extensions of this idea have been applied to systems having separable environmental and internal couplings~\cite{nicoletti2021mutual,nicoletti2022mutual}, and to investigate interference effects arising from overlapping interactions ~\cite{nicoletti2024information}. However, mutual information is a symmetric and time-independent measure for a stationary process and does not capture the directional or dynamical properties, limiting its ability to reveal causal driving or flow of information—both of which are central in understanding nonequilibrium functionality.

Directionality becomes especially important in biological systems, where the ability to channel and route information is often essential for function. For example, signal processing in gene regulatory networks relies not just on correlations but on well-defined causal structures~\cite{tyson2001network,tkavcik2008information}. To address this, time-delayed information measures, such as time-delayed mutual information~\cite{vastano1988information,kirst2016dynamic,li2018causal} and transfer entropy~\cite{schreiber2000measuring,kaiser2002information,hempel2024simple}, have emerged as powerful tools for uncovering directed dependencies from time-series data. These measures enable us to identify driving variables, causal directionality, and the locus of nonequilibrium activity — even in the presence of complex interactions~\cite{sanchez2002nonequilibrium,lozano2022information,loos2023long,ito2013information,ito2015maxwell,samsuzzaman2025flow,sagawa2013role,chetrite2019information,spinney2016transfer}.\\

\begin{figure*}
    \centering
    \includegraphics[width=0.9\linewidth]{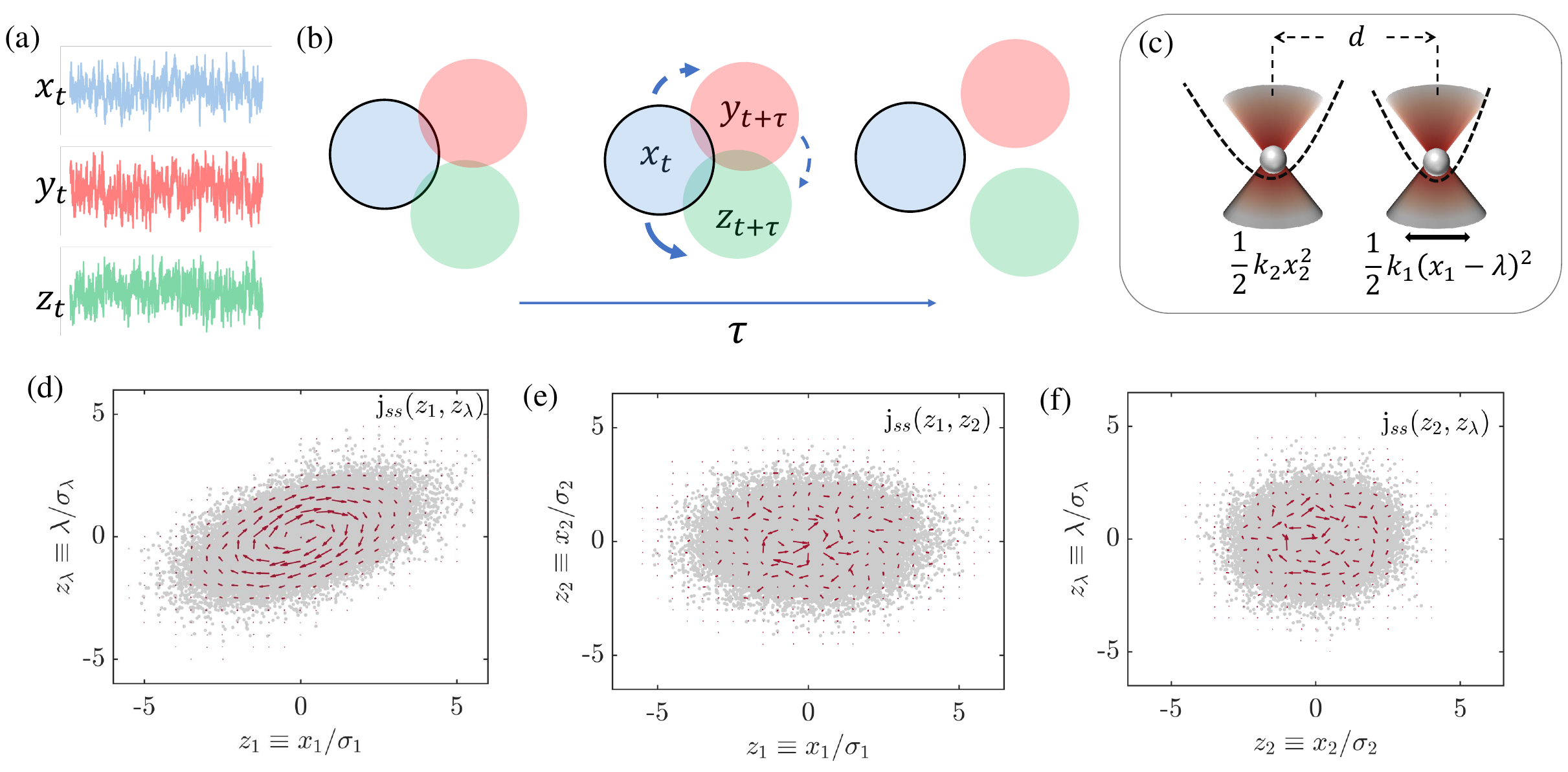}
    \caption{(a) Spatio-temporal trajectories corresponding to the three interacting variables $x$, $y$ and $z$ of a generic system. (b) 
    The mutual dependence of these variables is also a function of the lag time $\tau$. The extent of overlap at intermediate $\tau$ is informative of the causal directionality indicated by the arrows.
    We can establish such causal structure between the variables by estimating information theoretic measures~\cite{vastano1988information,schreiber2000measuring}. (c) Schematic diagram of the experimental system consisting of two hydrodynamically coupled microparticles trapped in two optical traps separated by a distance $d$. The mean position of the trap with stiffness constant $k_1$ is modulated by the OU noise ($\lambda(t)$), while the other trap ( with stiffness constant $k_2$) remains fixed. The fluctuating positions of the particles with
    respect to the center of each optical trap are denoted by $x_1(t)$ and $x_2(t)$. (d)-(f): Probability currents~\cite{li2019quantifying, das2024irreversibility} (shown as maroon arrows) computed from experimentally recorded spatio-temporal trajectories in every bivariate subspace of our system in the presence of external \textit{active} drive. The dynamical variables $(x_1,x_2,\lambda)$ are normalised to make them dimensionless. Here, $\sigma_{1,2}=\sqrt{k_BT/k_{1,2}}$ and  $\sigma_\lambda = \sqrt{D_e/\tau_e}$.}
    \label{fig:gen_dia}
\end{figure*}

To explore these ideas in a controlled setting, we consider a well-characterized mesoscopic system: two colloidal particles hydrodynamically coupled and confined in separate harmonic traps. While the hydrodynamic interaction couples their dynamics, the particles remain statistically independent in equilibrium~\cite{meiners1999direct}. However, when one of the traps is driven by exponentially correlated noise (modeled via an Ornstein-Uhlenbeck process), the system is driven out of equilibrium, and directional effects emerge. This kind of active noise can be seen as a stand-in for biochemical or environmental fluctuations with memory—akin to stimuli from an active bath~\cite{maggi2014generalized, dabelow2019irreversibility, majhi2025decoding}.

In a recent study, we demonstrated a rather counterintuitive interplay between irreversibility (measured by the total entropy production rate~\cite{seifert2005entropy,landi2021irreversible,manikandan2020inferring,manikandan2021quantitative,das2022inferring, das2025localizing}) and interactions in this system~\cite{das2024irreversibility}. Specifically, we showed that increasing the interaction strength by reducing the distance between particles lowers the total irreversibility of the entire system. However, when focusing on coarse-grained subspaces that include only the particles (excluding the active drive), the trend reverses: the measured irreversibility increases with interaction strength. This presents an apparent conundrum - would the degree of coarse-graining performed dictate the nature of irreversibility?
Here, we use information-theoretic measures as tools to pinpoint the driving node, reconstruct the causal structure between subsystems, and quantify how hydrodynamic interactions shape both the directionality of information flow and, crucially, the emergence of this counterintuitive behavior. 

The article is structured as follows: in section~\ref{sec:theory_info}, we present a short overview of the theoretical framework of information theoretic measures that we are interested in computing. Then, we describe our experimental system in brief in section~\ref{sec:theory_model}. In section~\ref{sec:info_theory_analysis}, we estimate information theoretic measures from experimental and numerical trajectories and discuss our observations. Finally, we conclude in Section~\ref{sec: conclusion}.
\section{Theoretical formalism: Information theoretic measures}
\label{sec:theory_info}
Our focus of attention in this work lies in understanding the effect of interaction shaping the mutual dependence and, more importantly, the causal directionality of interactions between different degrees of freedom of a stochastic system. To comprehend the basic theoretical formalism we have adapted, consider -- for simplicity -- a generic system in a steady state with three interacting subsystems or variables $x$, $y$ and $z$ with nearly identical spatio-temporal trajectories (Fig.~\ref{fig:gen_dia}(a)) -- the extension to higher dimensional systems is straightforward. 
Recent studies have shown that information theoretic measures can be utilised to inspect both plausible mutual dependencies and causal relationships between these variables, in a data-driven manner (Fig.~\ref{fig:gen_dia}(b)).  We begin with introducing the measures that we consider in this study. \\

\textit{Mutual information.-} 
The mutual information (MI, in \textit{bits} unit) between any two continuous stochastic variables $x(t)$ and $y(t)$ is defined as~\cite{shannon1948mathematical,cover2005elements},
\begin{equation}
    \label{eq:MI_def}
    MI_{xy} = \int dxdy\ P_{xy}(x,y) \log_2 \frac{P_{xy}(x,y)}{P_x(x) P_y(y)},
\end{equation}
which is the Kullback Leibler (KL) divergence between the joint distribution $P_{xy}(x,y)$ and the product of corresponding marginals $P_x(x)$ and $P_y(y)$. By definition, $MI_{xy}\geq0$ and $MI_{xy} = 0$ only when the variables are independent. Recently, mutual information has been widely used in stochastic processes to understand the effects of complex environments~\cite{nicoletti2021mutual,nicoletti2022mutual} and the role of activity~\cite{nicoletti2024information,nicoletti2024tuning}, even in several biological phenomena such as cell signaling~\cite {selimkhanov2014accurate,bauer2021trading,kramar2024single} and neuronal signal processing~\cite{quian2009extracting,barzon2025excitation}. It has also been used as a measure to quantify the distance between two distributions in non-equilibrium transformations~\cite{lapolla2020faster,ibanez2024heating,rose2024role,ManikandanEq}.  
In this most basic form, by construction, $MI_{xy}$ is a symmetric and time-independent quantity, thereby incapable of detecting any causal characteristics or dynamics of information transfer from one state to another. In this context, two other information theoretic measures -- time-delayed mutual information and transfer entropy --  have been proposed. We now turn our attention to these entities.\\

\textit{Time-delayed mutual information.-}
 The time-delayed mutual information from the variable $x$ to $y$ defined as~\cite{vastano1988information,kirst2016dynamic,li2018causal}, 
 
\begin{align}
\begin{split}
    MI_{x \rightarrow y} (\tau) \\
    &\hspace{-50pt}= \int dx_t dy_{t+\tau}\ P_{xy}(x_t,y_{t+\tau}) \log_2 \frac{P_{xy}(x_t,y_{t+\tau})}{P_x(x_t) P_y(y_{t+\tau})}, \\
    &\hspace{-50pt}\equiv \int dx_t dy_{t+\tau}\ P_{xy}(x_t,y_{t+\tau}) \log_2 \frac{P_{xy}(y_{t+\tau}|x_t)}{P_y(y_{t+\tau})},
\end{split}
\label{eq:MI_def_timelagged}
\end{align}

where $\tau$ denotes the finite time lag between two variables. This quantity essentially measures mutual information between two segments of trajectories that are time-delayed with respect to the other, such as $x_t \equiv (x^1,x^2,...,x^{n-m})$ and $y_{t+\tau} \equiv (y^{1+m}, y^{2+m}, ..., y^n)$. Here, $n$ denotes the block of trajectory length up to a fixed time $t$ and $m = \tau/\Delta t$,  denotes the number of points corresponding to a time delay  $\tau$. Note that we compute the mutual information between these finite segments, not between entire stochastic trajectories in the path-wise sense~\cite{reinhardt2023path}. Unlike steady-state mutual information, its time-delayed version is asymmetric in general, {\it i.e.}, $MI_{x \rightarrow y} (\tau) 
 \neq MI_{y \rightarrow x} (\tau) $. Furthermore, since its structure is explicitly lag time dependent -- it can be leveraged to detect the effective directional information transfer in $(x,y)$ phase space. A nonzero value of the time-delayed mutual information as a function of $\tau$ indicates causal interactions between two variables. For a generic stationary process, it can be shown that $MI_{x\rightarrow y}(-\tau) = MI_{y \rightarrow x} (\tau)$,  suggesting to the fact that while $MI_{x \rightarrow y}(\tau)$ quantifies how the present $x_{t}$ predicts the future $y_{t+\tau}$, for a negative lag, this simply swaps to how the present  $y_{t}$ predicts the future  $x_{t+\tau}$.
 The sign of the time lag, $\tau^{\max}$, at which this quantity reaches its peak magnitude (global maximum) can then be used to infer the direction of information flow~\cite{li2018causal}. Hence, a positive $\tau_{max}$ in Eq.~\eqref{eq:MI_def_timelagged} indicates that $y$ shares maximum information with the past of $x$, which implies that $x$ drives $y$. Also, if $(MI_{x \rightarrow y}(\tau) > MI_{y \rightarrow x}(\tau) )$, it can be statistically considered that the information is predominantly shared from the variable (or process) $x$ to the variable $y$ and not the other way around ~\cite{kirst2016dynamic}.

However, it is important to point out that time-delayed mutual information only measures the difference between $P_{xy}(y_{t+\tau}|x_t)$ and $P_y(y_{t+\tau})$ -- which quantifies the reduction in uncertainty regarding the future of $y$, thereby incorporating the history of $x$. By definition, it does not exclude the effects induced by its own ($y$) history~\cite{kirst2016dynamic}.  
To remove this effect, another time-dependent measure - the transfer entropy -  was introduced by T. Schreiber in Ref.~\cite{schreiber2000measuring}.\\
\textit{Transfer entropy.-} The transfer entropy is defined as~\cite{schreiber2000measuring, kaiser2002information,hempel2024simple},
\begin{align}
\label{eq:TE}
\begin{split}
    TE_{x\rightarrow y}(\tau)&\\
   &\hspace{-50pt}=\int dx_tdy_tdy_{t+\tau}\ P_{xy}(y_{t+\tau}, y_t,x_t) \log_2\frac{P_{xy}(y_{t+\tau}|y_t, x_t)}{P_y(y_{t+\tau}|y_t)}. 
    \end{split}
\end{align}
This quantity basically measures the reduction of uncertainty regarding the present value of $y$ by knowing the history of $x$, given that the history of $y$ is known as well. Unlike time-delayed mutual information, transfer entropy is argued to exclude the variable's own history and thereby captures purely exchanged information between the subsystems. This measure is also asymmetric by construction as $ TE_{x\rightarrow y}(\tau) \neq TE_{y\rightarrow x}(\tau)$, which is suggestive of a prominent direction of information flow. Furthermore, if $TE_{x\rightarrow y}(\tau)  \gg TE_{y\rightarrow x}(\tau) $, the evidence of (causal) directional influence from the process $x$ to the process $y$ can be statistically ensured~\cite{hempel2024simple}.
Note that, for negative time-lags, $TE_{x \rightarrow y}(-\tau)$ sets conditions on the future of the target, so it does not follow the simple sign-flip symmetry of time-delayed mutual information. As a result, $TE_{x \rightarrow y}(-\tau)$ does not generally equal 
$TE_{y \rightarrow x}(\tau)$. It is also expected that the standard positive-lag TE remains more directly meaningful for causal interpretation.

Both of these quantities are used in various domains to understand the causal connection and structure of connectivity between different variables~\cite{lozano2022information, ito2015maxwell, kirst2016dynamic,samsuzzaman2025flow}. For example, time-delayed mutual information was used to explore spatiotemporal information transport in physical systems~\cite{vastano1988information} and to infer causal interactions in neural network systems~\cite{endo2015delayed}, and chaotic systems~\cite{ho2003information}. On the other hand, transfer entropy is typically employed to understand the information flows within gene-regulatory networks~\cite{tyson2001network,tkavcik2008information}, biochemical signaling~\cite {ito2015maxwell} and various other systems characterized by long memory~\cite{staniek2008symbolic,dickten2014identifying}. More recently, such measures have also been utilised to reconstruct a directed graph for a system with many interacting dynamical variables that are spatially separated~\cite{hempel2024simple}. In an alternate context, the rates of such quantities are also used to derive a generalised second law of information thermodynamics for systems without bipartite structure~\cite{chetrite2019information}. Moreover, the connection between such information theoretic measure (particularly transfer entropy) and the response in a spatially asymmetric extended system is also explored~\cite{sarra2021response,lucente2024conceptual}.

Given the unique advantages of information-theoretic measures, we seek to apply these tools to our experimental non-equilibrium system, where hydrodynamic interactions are prevalent. 
The system consists of two colloidal particles in two separate harmonic traps, placed in very close separation so that hydrodynamic interactions naturally arise (Fig.~\ref{fig:gen_dia}(c)). In addition, one of the particles is driven by an external stochastic protocol, which makes the overall process non-equilibrium, characterised by probability fluxes~\cite{li2019quantifying,das2024irreversibility} in different phase spaces (Fig.~\ref{fig:gen_dia}(d)-(f)). 
Unlike conventional interactions of conservative origin, hydrodynamic interactions are dissipative and can, in certain cases, intrinsically link both deterministic and stochastic forces acting on the interacting degrees of freedom~\cite{meiners1999direct}. This leads to an emergent \textit{non-multipartite}  structure~\cite{leighton2024jensen} in the dynamics. Interestingly, it turns out that such a dissipative and multi-partite structure significantly influences the irreversibility and entropy production of this system. 

\begin{figure*}
    \centering
\includegraphics[width=0.9\linewidth]{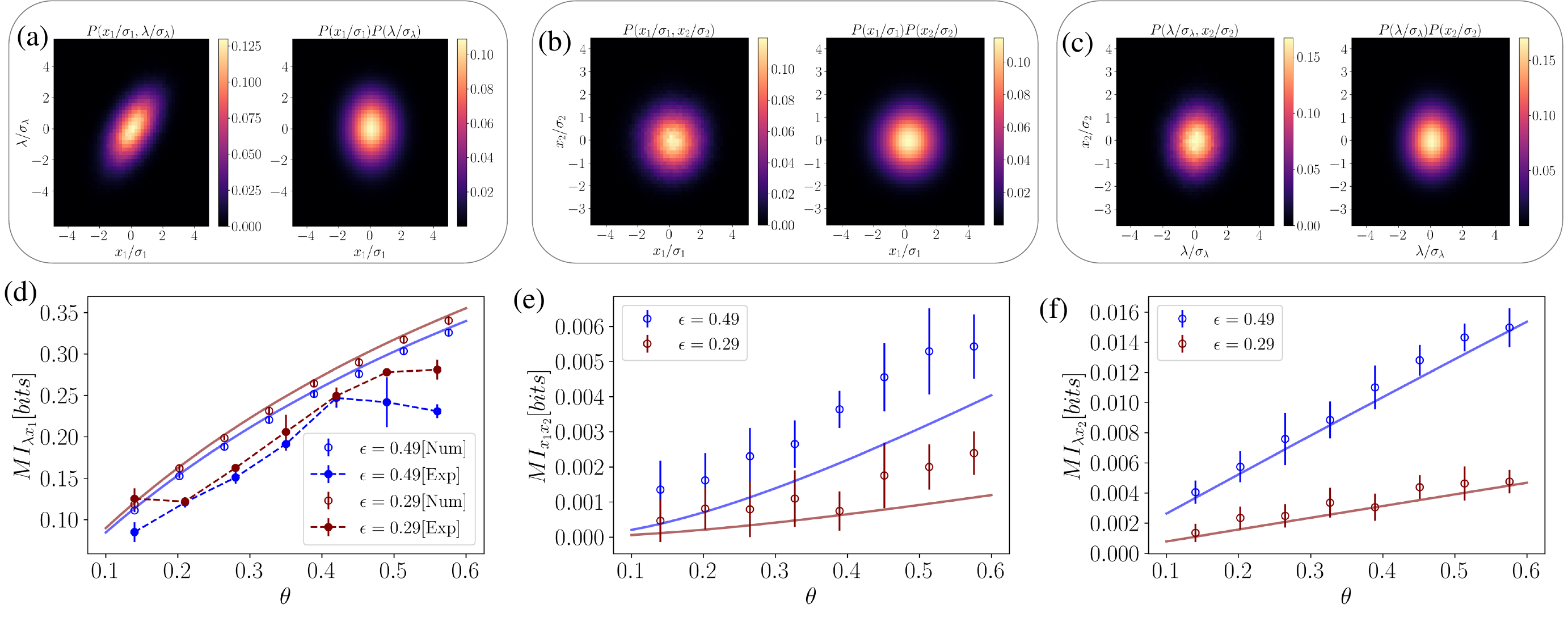}
    \caption{\textbf{System with hydrodynamic interaction:} (a)-(c): Comparison between the joint probability distribution functions of any two dynamical variables of the system (such as $P(\lambda,x_1)$, $P(x_1,x_2)$ and $P(\lambda,x_2)$) and product of the corresponding marginals (such as $P(\lambda)P(x_1)$, $P(x_1)P(x_2)$ and $P(\lambda)P(x_2)$) are shown for a particular nonequilibrium strength, $\theta = 0.56$ and hydrodynamic interaction strength, $\epsilon=0.49$.  The dynamical variables are normalised to make them dimensionless. Here, $\sigma_{1,2}=\sqrt{k_BT/k_{1,2}}$ and  $\sigma_\lambda = \sqrt{D_e/\tau_e}$. (d)-(f): Mutual information between different dynamical variables at steady-state are estimated for different nonequilibrium noise strengths ($\theta$) at two different interaction strengths ($\epsilon$). Here, the `open' symbols denote the mutual information estimated from the numerical trajectories and the `filled' symbols denote the same estimated from the experimental trajectories. $MI_{x_1 x_2}$ and $MI_{\lambda x_2}$ are estimated only from the numerical trajectories as the expected values (for the regimes of experimental parameters) are too low and may get overwhelmed by the noise level of the experiment. The solid lines represent corresponding analytical estimations. }
    \label{fig:PMIf}
\end{figure*}

In previous work, we showed that, rather counterintuitively, increasing interaction strength (by bringing the two colloidal particles closer to each other) reduces overall irreversibility and entropy production in this process, yet at the level of coarse-grained subspaces -- comprising only the particles -- this trend reverses~\cite{das2024irreversibility}. 
We were able to verify this trend over all possible ranges of parameter values for this system, both experimentally and with exact analytic solutions of the corresponding Langevin model.
This raises a fundamental question: What underlying dynamical dependencies drive these opposing trends in entropy production at different levels of description, and how do they shape the spatiotemporal structure of irreversibility in the system?

Here, we address this question by employing the aforementioned information-theoretic tools, on both the experimental and numerical data of this system. We demonstrate that these tools systematically dissect the role of hydrodynamic interactions in establishing dependencies between dynamical variables across all relevant phase planes and then go on to show how these dependencies evolve over time such that causal connections emerge between those. Crucially, they capture the apparent contradiction in irreversibility trends across different levels of description, providing a coherent framework for understanding nonequilibrium behavior in hydrodynamically coupled systems. We also discuss the interplay between hydrodynamic interactions and the strength of external driving in shaping the information flow in the full phase space.

\section{Experimental Model}
\label{sec:theory_model}
Our system consists of two identical Brownian particles trapped in two separate harmonic confinements at close separation in a viscous fluid.
Experimentally, the system is realised with the optical tweezers created by tight-focusing two perpendicularly polarised Gaussian beams emanating from two solid-state lasers of wavelength $1064~ nm$. To manifest a non-equilibrium scenario, the mean position of one of the traps is modulated by a correlated noise ($\lambda(t)$) with correlation timescale $\tau_e$ -- following an Ornstein-Uhlenbeck process such that $\langle\lambda(t)\lambda(t^\prime)\rangle = (D_e/\tau_e)\exp(-(t-t^\prime)/\tau_e)$~\cite{das2023enhanced}. One of the beams is passed through an acousto-optic modulator so as to add an external OU noise. Then the position fluctuations of both particles are separately recorded at the rate of $10~kHz$ for $50s$ -- using a `balanced-detector-system' consisting of high-gain photo-diodes. Further details about the experiment can be found in Ref.~\cite{das2024irreversibility}.

The dynamics of the microparticles, which are at close separation, is coupled via hydrodynamic interaction of strength $\epsilon = (3a/2d) - (a/d)^3$, where $d$ is the separation between the mean positions of the confinements and $a$ denotes the radius of the particle $(a = 1.5\mu m)$. In some ways, this system can be mirrored as a biochemical channel with two nodes, where one of the nodes has access to additional \textit{active} fluctuations acting as an external signal~\cite{hahn2023dynamical}. The schematic of the system is illustrated in Fig.~\ref{fig:gen_dia}(c).  Theoretically, the system with the particles trapped in optical potentials with different stiffness constants $k_1$ and $k_2$  can be expressed with coupled, linear \textit{Langevin} equations~\cite{das2024irreversibility}: 
\begin{widetext}
    \begin{equation}
   \label{eq:ap_dyn_neq1}
    \begin{split}
    \dot{x}_1(t) &= \mathcal{H}_{11}[-k_1 (x_1(t) - \lambda (t)) + \xi_1(t)] + \mathcal{H}_{12}[-k_2 x_2(t) + \xi_2(t)]\\
     \dot{x}_2(t) &= \mathcal{H}_{21}[-k_1 (x_1(t) - \lambda (t)) + \xi_1(t)] + \mathcal{H}_{22}[-k_2 x_2(t) + \xi_2(t)]\\
     \dot{\lambda}(t) &= -\lambda(t)/\tau_e+\sqrt{2D_e}/\tau_e\ \eta_3(t).
    \end{split}
\end{equation}
\end{widetext}
Here, $\mathcal{H}_{ij}$ $(i,j = 1,2)$ are the constant elements of a hydrodynamic coupling tensor of the form $\mathbf{\mathcal{H}} = \begin{pmatrix}
1/\gamma & \epsilon/\gamma\\
\epsilon/\gamma & 1/\gamma\\
\end{pmatrix}$ --  considering that the displacements of the particles (measured with respect to the centre of the corresponding trap) -- denoted as ($x_1(t)$, $x_2(t)$) are small compared to the mean separation between the traps ($d$) and $\gamma \equiv 6\pi\eta a$ is the viscous drag coefficient of the medium~\cite{berut2016stationary}. The terms $\xi_1(t)$ and $\xi_2(t)$ are delta-correlated random Brownian forces such that $\langle \xi_i(t)\rangle =0,\ \langle \xi_i(t)\xi_j(t)\rangle = 2 k_B T (\mathcal{H})^{-1}_{ij} \delta(t-t^\prime)$. $k_B$ is Boltzmann's constant. Additionally, $\langle \eta_3(t)\rangle = 0, \langle \eta_3(t)\eta_3(t^\prime)\rangle = \delta(t-t')$ and $\langle\xi_1(t)\eta_3(t')\rangle = \langle\xi_2(t)\eta_3(t')\rangle =0$. Eq.\eqref{eq:ap_dyn_neq1} can be rewritten as: $\mathbf{\dot{x}}(t) = -\mathbf{F\cdot x}(t) + \boldsymbol{\xi}(t)$ with $\mathbf{x}(t) = [x_1(t),x_2(t),\lambda(t)]^T$ and $\boldsymbol{\xi}(t) = [(\xi_1(t) + \epsilon\xi_2(t))/\gamma, (\epsilon\xi_1(t) + \xi_2(t))/\gamma,\eta_3(t)]^T \equiv [\eta_1(t), \eta_2(t), \eta_3(t)]^T$ such that $\langle\boldsymbol{\xi}(t):\boldsymbol{\xi}(t^\prime)\rangle = 2\mathbf{D}\delta(t-t^\prime)$. The interaction matrix ($\mathbf{F}$) and the diffusion matrix ($\mathbf{D}$) are of the forms:
\begin{equation}
\label{eq:drift_diffusive}
    \mathbf{F} = \begin{pmatrix}
        1/\tau_1 & \epsilon/\tau_2 & -1/\tau_1\\
        \epsilon/\tau_1 & 1/\tau_2  & -\epsilon/\tau_1\\
        0 & 0 & 1/\alpha\tau_1
    \end{pmatrix}, \
    \mathbf{D} = \begin{pmatrix}
        D_0 & \epsilon D_0 & 0\\
        \epsilon D_0 & D_0 & 0\\
        0 & 0 & \theta D_0
    \end{pmatrix}, 
\end{equation}
where, $\tau_{1,2} = \gamma/k_{1,2}, \alpha = \tau_e/\tau_1, D_0 = k_B T/\gamma $ and $\theta = (D_e/\tau_e^2)/D_0$. In the experiment, $D_0 \sim 0.16~\mu m^2/s$, $\tau_1 \sim 1.5~ms$, $\tau_2 \sim 2.1~ms$ and $\tau_e = 4~ms$ are fixed such that $\alpha\sim2.7$. We vary the nonequilibrium strength by $D_e$ such that $\theta$ varies from $0.14-0.56$ at two different separations of the particles, corresponding to the hydrodynamic interaction strengths of $\epsilon = 0.49$ and $\epsilon=0.29$.  Note that the diffusion matrix at any separations is non-diagonal, which is a characteristic feature in the case of hydrodynamic interactions as it encodes the fact that the noise acting on one particle is correlated with the motion of the other due to fluid-mediated couplings. The steady-state distribution of the system over the phase space defined by ($x_1,x_2,\lambda$) will be a multivariate Gaussian such that, $P(x_1,x_2,\lambda)\sim \mathcal{N}(0, \mathbf{C}) \sim \exp(-\frac{1}{2}\mathbf{x}^T \mathbf{C}\mathbf{x})$ and the covariance matrix ($\mathbf{C}$) can be computed by solving the Lyapunov equation: $\mathbf{FC} + \mathbf{C}\mathbf{F}^T = 2\mathbf{D}$. Interestingly, the steady-state distribution of the system at marginal bivariate subspaces (i.e. $(x_1,x_2)$, $(x_1,\lambda)$, $(x_2,\lambda)$) will also be Gaussian. For example, $P(x_1,x_2)\sim \mathcal{N}(0, \mathbf{C}_{x_1x_2})$, where $\mathbf{C}_{x_1x_2}$ is a submatrix of $\mathbf{C}$ whose rows and columns correspond to $x_1$ and $x_2$. The elements of the steady-state covariance matrix ($\mathbf{C}$) are explicitly shown in \textit{Appendix}~\ref{ap:cov_matrix}.
\section{Results and Discussions}
\label{sec:info_theory_analysis}

\begin{figure*}
    \centering
    \includegraphics[width=0.9\linewidth]{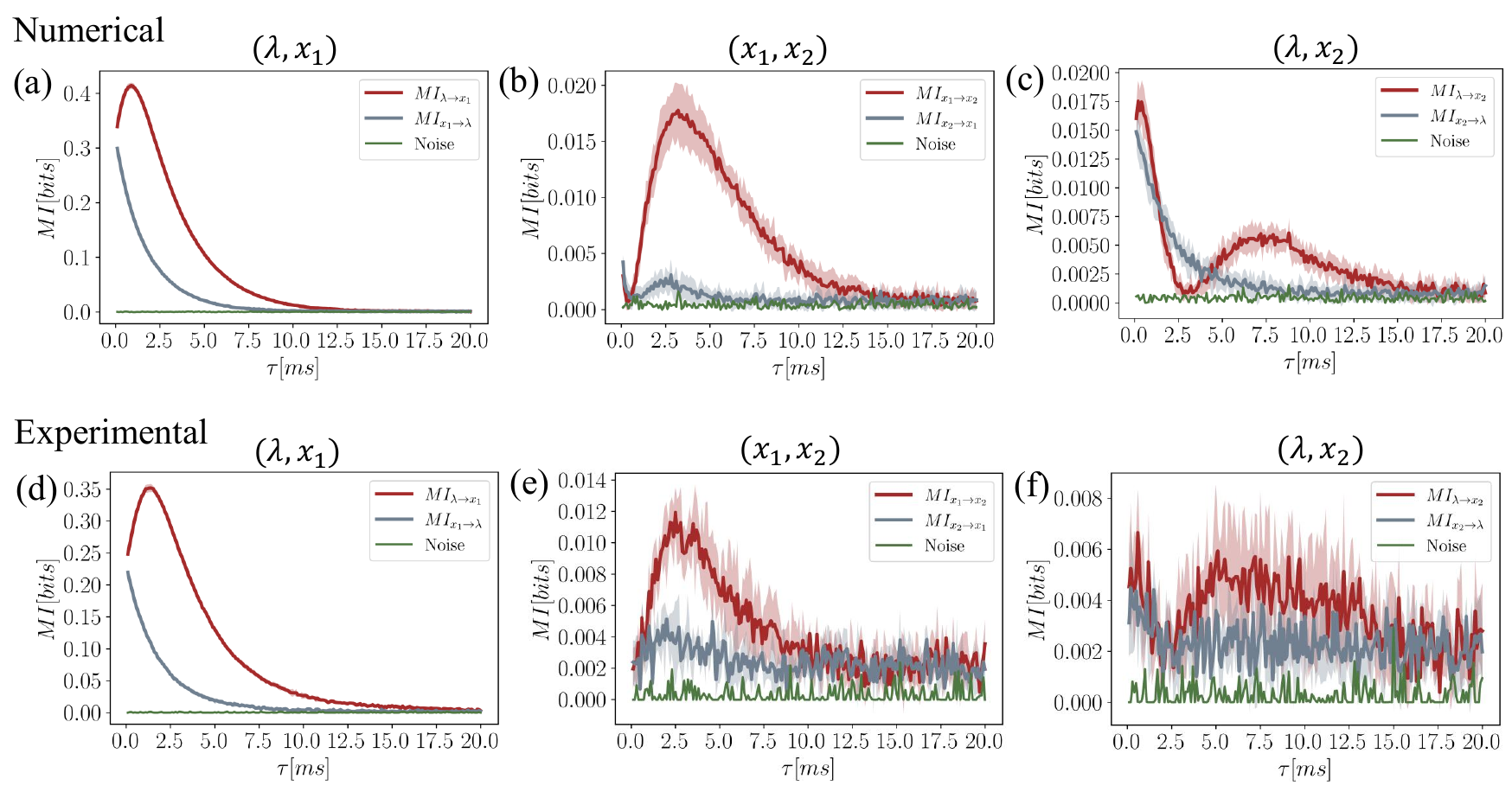}
    \caption{Estimated time-delayed mutual information between different dynamic variables of the system with $\theta = 0.56$ and $\epsilon = 0.49$ - in (a) \& (d) $(\lambda,x_1)$, (b) \& (e) $(x_1,x_2)$ and (c) \& (f) $(\lambda,x_2)$ spaces.
    The shaded regions in all plots are the standard deviations of the estimations performed over multiple trajectories and the solid lines are the mean of corresponding estimations. The `green' line in each plot is the threshold value of mutual information between pairs of dynamical variables above which the value of mutual information can be considered as `significant'. It is obtained by calculating mutual information from the randomly shuffled trajectories.}
    \label{fig:MI_time}
\end{figure*}

 We begin by estimating $MI_{\lambda x_1}$, $MI_{\lambda x_2}$ and $MI_{x_1x_2}$ as we aim to investigate how hydrodynamic interactions affect the interaction of the individual particles with the external drive as well as the coupling between the trapped particles under an external nonequilibrium drive. Due to the Gaussian features of the $P_{x_1x_2}$ and its marginals, the mutual information between $x_1$ and $x_2$ can be analytically computed as~\cite{nicoletti2024information,nicoletti2024tuning},
\begin{equation}
\label{eq: MIx1x2}
    MI_{x_1x_2} = \frac{1}{2}\log_2\frac{c_{x_1x_1} c_{x_2x_2}}{\det\mathbf{C}_{x_1x_2}},
\end{equation}
where $c_{x_1x_1}$ and $c_{x_2x_2}$ are the diagonal entries of the corresponding submatrix $\mathbf{C}_{x_1x_2}$ of the covariance matrix $\mathbf{C}$.  Similarly, the mutual information between the external drive $\lambda$ and the driven particle can be estimated as,
\begin{equation}
\label{eq: MIlambdax1}
    MI_{\lambda x_1} = \frac{1}{2}\log_2\frac{c_{\lambda \lambda}c_{x_1x_1} }{\det \mathbf{C}_{\lambda x_1}},
\end{equation}
using the submatrix $\mathbf{C}_{\lambda x_1}$ and its diagonal entries. On the other hand, $MI_{\lambda x_2}$ can also be similarly calculated using the submatrix $\mathbf{C}_{\lambda x_2}$. The analytical expressions of these quantities are explicitly shown in \textit{Appendix}~\ref{ap:MI}. Moreover, the signature of non-zero mutual information will be encoded in the characteristic differences between the joint probability distribution of two variables and the product of the corresponding marginals. As we compare different joint probability distributions and the products of the marginals in Fig.~\ref{fig:PMIf}(a)-(c), it is clear that mutual information between $\lambda$ and $x_1$ will be significantly higher than the mutual information between $x_1$ and $x_2$ or $\lambda$ and $x_2$. This is consistent with the fact that the active drive ($\lambda$) is directly applied to the mean position of the first trap so that the position fluctuations of the first particle ($x_1$) get strongly affected by the drive, which leads to significantly higher dependence between them. On the other hand,  both the interparticle interaction and the connection between the second particle ($x_2$) and the drive ($\lambda$) are mediated by hydrodynamic interaction which leads to relatively weaker dependence.

We then use a $\textit{k}$-nearest neighbour (kNN) estimator~\cite{kraskov2004estimating} to compute mutual information between different variables from the experimental and numerical trajectories. It is known that 
$\textit{k}$-nearest neighbour estimator relies on non-parametric methods to compute mutual information in a data-driven manner so that it naturally adapts to the local structure of the data distribution irrespective of the underlying probability distributions~\cite{ross2014mutual}. 

Interestingly, we find that the steady state mutual information between the driven particle ($x_1$) and the external drive ($\lambda$) is less at higher hydrodynamic interaction strength ($\epsilon$) at any strengths of the nonequilibrium drive ($\theta$)[Fig.~\ref{fig:PMIf}(d)]. 
On the other hand, the mutual information between other dynamical variables ($MI_{x_1x_2}$ and $MI_{\lambda x_2}$) is indeed enhanced in the case of strong hydrodynamic interaction for any strength of the external drive, as shown in Fig.~\ref{fig:PMIf}(e) and Fig.~\ref{fig:PMIf}(f). 
These observations are nontrivial and suggest that the interactions with the other particle reduce the effect of the non-equilibrium drive that the first particle experiences as compared to the non-interacting limit. We can indeed trace back the decrease in the total entropy production rate to this observation. Additionally, the enhancement of mutual information between the particles ($MI_{x_{1}x_{2}}$) at higher interaction strengths also corroborates our previous observation related to the trend of irreversibility in the coarse-grained phase space~\cite{das2024irreversibility}.  

\textit{Time-delayed informational measures.-} 
We show that the mutual information successfully captures the effect of the hydrodynamic interactions in building the mutual dependencies between the dynamical variables of each phase plane. Still, it is not enough to pinpoint the dominant variable that drives the dynamics of the other in a phase plane, which means that the dynamics or direction of information flow remains unknown. Since directional information flow indicates the direction of interactions between local subsystems -- which may have impacts on the overall dynamical and thermodynamical properties of the system -- we next compute time-delayed mutual information (as defined in Eq.~\eqref{eq:MI_def_timelagged}) between different dynamical variables to understand the possible directional dependencies in our system. 

Similar to the time-independent MI, we compute mutual information between two variables using the \textit{k}-nearest neighbour estimator~\cite{kraskov2004estimating}, but now with a pair of variables where one among them lags behind the other with a lag time $\tau$. In other words, we compute different combinations of mutual information from the time-delayed experimental and numerical trajectories. 
Specifically, we estimate the time-delayed mutual information parameters $MI_{\lambda\rightarrow x_1}(\tau) $, $MI_{x_1\rightarrow x_2}(\tau) $
and $MI_{\lambda\rightarrow x_2}(\tau) $ along with their reverse counterparts to determine the directionality of information flow.

Interestingly, this quantity turns out to be asymmetric under the change of arguments. 
As we show in Fig.~\ref{fig:MI_time}, for  nonequilibrium strength $\theta = 0.56$ and hydrodynamic interaction strength $\epsilon = 0.49$,  $MI_{\lambda\rightarrow x_1}(\tau)  > MI_{x_1\rightarrow \lambda}(\tau) $
, $MI_{x_1\rightarrow x_2}(\tau)  > MI_{x_2\rightarrow x_1}(\tau) $, and $MI_{\lambda\rightarrow x_2}(\tau)  > MI_{x_2\rightarrow \lambda}(\tau) $ which strongly indicates the preferred direction of information flow.  Additionally, $MI_{\lambda\rightarrow x_1}(\tau) $, $MI_{x_1\rightarrow x_2}(\tau) $
and $MI_{\lambda\rightarrow x_2}(\tau) $ have global maxima at different time lags. While two of the reverse counterparts, $MI_{x_1\rightarrow \lambda}(\tau) $ and $MI_{x_2\rightarrow \lambda}(\tau) $ do not possess any such maxima, interestingly, $MI_{x_2\rightarrow x_1}(\tau) $ shows a maxima at a particular time lag, similar to   $MI_{x_1\rightarrow x_2}(\tau) $. The presence of a maxima in $MI_{x_2\rightarrow x_1}(\tau) $ indicates that there is a possibility of information backflow from the fixed particle ($x_2$) to the driven one ($x_1$). On the other hand, the exponential-type decay that appears in $MI_{x_1\rightarrow \lambda}(\tau) $ and $MI_{x_2\rightarrow \lambda}(\tau) $ is due to the inherent nature of time-delayed mutual information, where it fails to exclude the effects induced due to the history of $\lambda$ (which is exponentially correlated), as can be understood from the second definition described in Eq.~\eqref{eq:MI_def_timelagged}. Moreover, in our system, it is physically impossible for information to flow from any of the particles to the external drive since the drive is independent of them. As a result, the information pathways in the presence of hydrodynamic interaction can be diagrammatically represented as,
\tikz[baseline=(x1.base), every node/.style={font=\footnotesize}, >=Stealth] {
  \node (lambda) {$\lambda$};
  \node (x1) [right=of lambda] {$x_1$};
  \node (x2) [right=of x1] {$x_2$};

  \draw[->,line width=1.4pt] (lambda) -- (x1);
  \draw[->,line width=1pt, bend left=10] (x1) to (x2); 
  \draw[->, line width=0.3pt, bend left=10] (x2) to (x1);
  \draw[->, line width=0.3pt, bend left=25] (lambda) to (x2);
}, for moderate nonequilibrium strengths (such as $\theta = 0.56$). However, if the nonequilibrium strength is enhanced such that $\theta \gg 1$ (which is beyond our experimental regime), the information pathways for our system will be fully unidirectional as,\tikz[baseline=(x1.base), every node/.style={font=\footnotesize}, >=Stealth] {
  \node (lambda) {$\lambda$};
  \node (x1) [right=of lambda] {$x_1$};
  \node (x2) [right=of x1] {$x_2$};

  \draw[->,line width=1.4pt] (lambda) -- (x1);
  \draw[->,line width=1pt] (x1) to (x2); 
  \draw[->, line width=0.3pt, bend left=20] (lambda) to (x2);
}.  
We further demonstrate this with numerical evidence in Fig.~\ref{fig:MI_TE_highTheta} of {\textit{Appendix}~\ref{ap:TE_MI_both}. Note that we present estimates of time-delayed mutual information using a kNN estimator in a fully data-driven manner. For this system, however, the same quantity can also be calculated analytically in terms of the two-point correlation function, since the dynamics are Gaussian. Consequently, the time lag at which the global maximum occurs for the time-delayed mutual information between any two variables corresponds to the time of maximum correlation or anti-correlation, e.g., $\langle x(t) y(t+\tau) \rangle$. The time-dependent cross-correlation functions between different dynamical variables are displayed in \textit{Appendix}~\ref{ap:ccf}. As further shown in \textit{Appendix}~\ref{ap:tdmi_analytics}, the analytical estimates of time-delayed mutual information reasonably match the numerical estimates.
All in all, this informational pathway is indeed a signature of \textit{hierarchical} interaction as the presence of the external active drive ($\lambda$)  is partially available to the fixed particle ($x_2$) through the hydrodynamic interaction.

 \begin{figure*}
    \centering
    \includegraphics[width=0.9\linewidth]{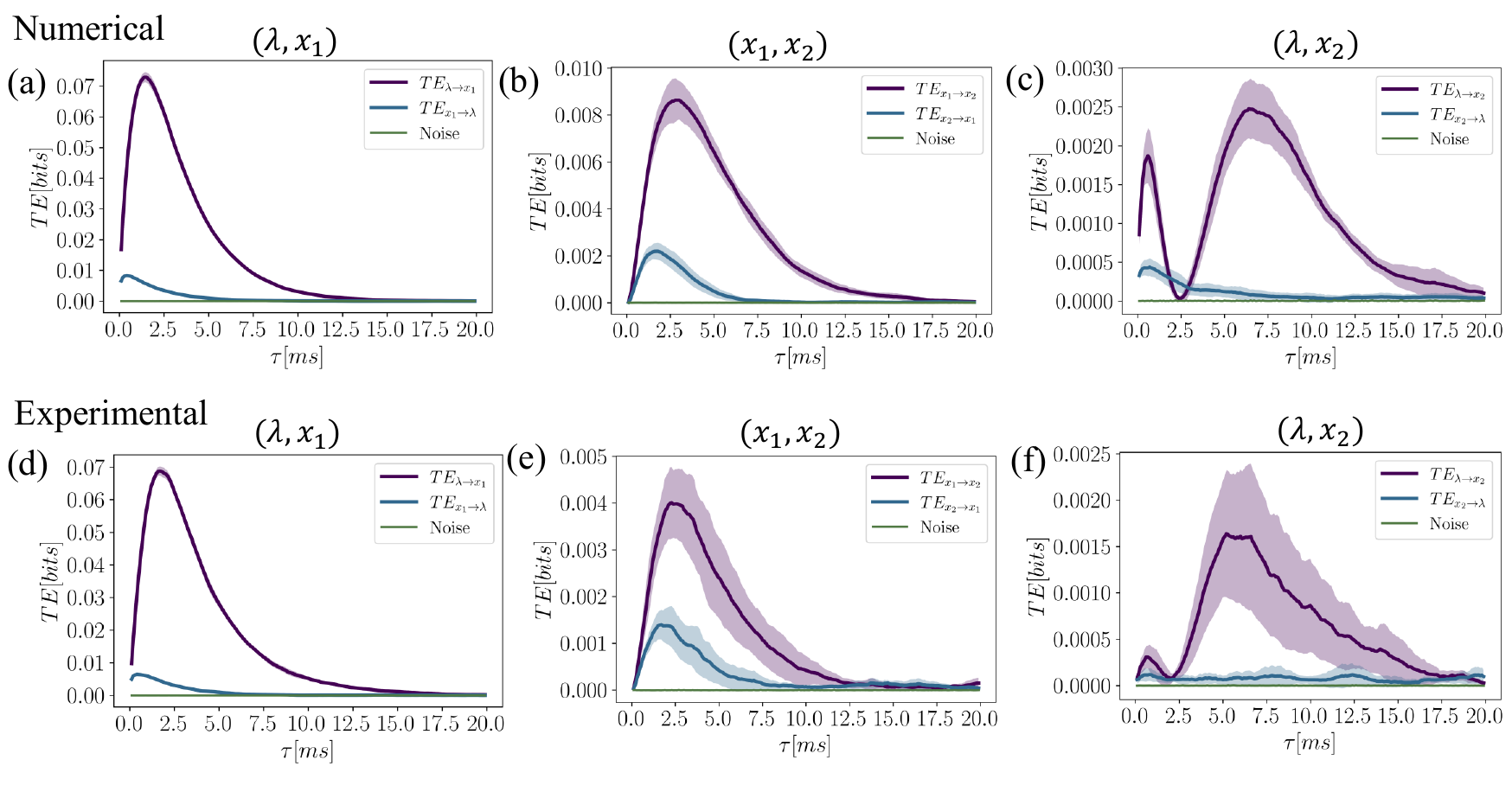}
    \caption{Estimated transfer entropies between different dynamic variables of the system with $\theta = 0.56$ and $\epsilon = 0.49$ - in (a) \& (d) $(\lambda,x_1)$, (b) \& (e) $(x_1,x_2)$ and (c) \& (f) $(\lambda,x_2)$ spaces. The shaded regions in all plots are the standard deviations of the estimations performed over multiple trajectories and solid lines are the mean of corresponding estimations. The `green' line in each plot is the threshold value of transfer entropy between pairs of dynamical variables above which the value of transfer entropy can be considered as `significant'. It is obtained by calculating transfer entropy from the randomly shuffled trajectories. }
    \label{fig:TE_time}
\end{figure*}

To establish this causal directionality further, we next estimate \textit{transfer entropy} (as defined in Eq.~\eqref{eq:TE}), which is capable of capturing only the exchanged information between two subsystems by excluding the history of the target.
To compute the transfer entropy from the experimental and numerical trajectories, we have used the recently proposed  Markov-state-model(MSM)-based estimation technique because of its simplicity and calculation efficiency~\cite{hempel2024simple, noteTE}. 

Similar to time-delayed mutual information, $TE_{\lambda\rightarrow x_1}(\tau) $, $TE_{x_1\rightarrow x_2}(\tau) $
and $TE_{\lambda\rightarrow x_2}(\tau) $ also possess global maxima at different time lags (shown in Fig.~\ref{fig:TE_time}) and $TE_{\lambda\rightarrow x_1}(\tau)  \gg TE_{x_1\rightarrow \lambda}(\tau) $, $TE_{\lambda\rightarrow x_2}(\tau)  \gg TE_{x_2\rightarrow \lambda}(\tau) $. Additionally, a global maximum is indeed present in $TE_{x_2\rightarrow x_1}(\tau) $.  The particular time lag at which the global maximum occurs for the transfer entropy between any two dynamical variables corresponds to the time of maximum correlation or anti-correlation between those two variables as well. Moreover, the appearance of two peaks in $TE_{\lambda\rightarrow x_2}(\tau) $ is also related to the non-intuitive behaviour of $\langle \lambda(t)x_2(t+\tau)\rangle$ - which possesses two local extrema at two different time-lags. It is worth noting that while the time-dependent cross-correlation function quantifies linear correlations between the dynamic variables, the time-delayed mutual information and the transfer entropy are capable of capturing both linear and nonlinear dependencies between different variables of a generic system, thereby revealing richer dynamical structure~\cite{li2018causal}.

\begin{figure*}
    \centering
    \includegraphics[width=0.95\linewidth]{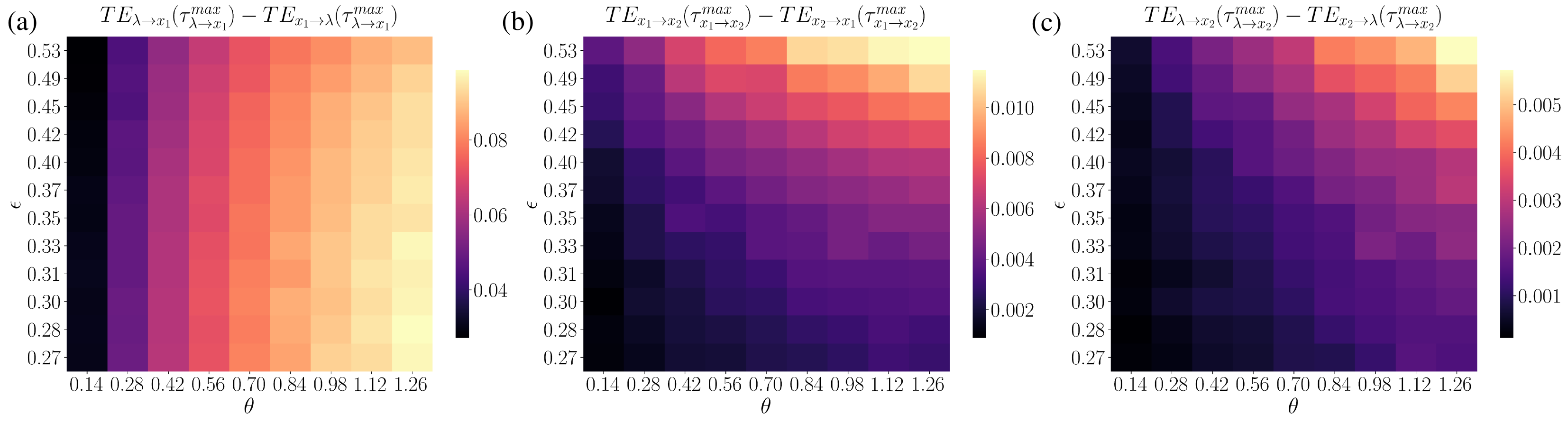}
    \caption{Asymmetry of transfer entropies as a function of $\theta$ and $\epsilon$,  estimated at the lag time corresponding to the maximum information transfer in (a) $(\lambda, x_1)$, (b) $(x_1, x_2)$, and $(\lambda, x_2)$ spaces, respectively. The values represent the mean of 10 independent estimations, each based on distinct numerical trajectories, for every pair of $\theta$ and $\epsilon$.  }
    \label{fig:TE_asy}
\end{figure*}

The overall features of transfer entropies between different dynamical variables indicate a unique direction of information flow that matches with the direction suggested by the characteristics of the time-delayed mutual information. Moreover, it can be further ascertained that in a moderate nonequilibrium scenario, the information pathways between hydrodynamically coupled particles are bi-directional, which can be made completely unidirectional by enhancing the strength of the nonequilibrium drive. 
Moreover, the asymmetry of the time-delayed quantities also depends on $\epsilon$ for a given $\theta$ - as shown in Fig.~\ref{fig:MI_TE_loweps} of {\textit{Appendix}~\ref{ap:TE_MI_both} -  with the time-delayed information measures estimated for $\theta = 0.56$ and $\epsilon = 0.29$. 

To further explore such dependencies on the system parameters, we introduce a measure of asymmetry in causal directionality for any generic $(x,y)$ phase space, defined as the difference in magnitudes of the transfer entropies estimated at the lag time corresponding to the maximum information transfer of $TE_{x\rightarrow y}$, i.e. $\mathcal{A}_{x\rightarrow y}(\tau^{max}_{x\rightarrow y}) = TE_{x\rightarrow y}(\tau^{max}_{x\rightarrow y}) - TE_{y\rightarrow x}(\tau^{max}_{x\rightarrow y})$.
We then compute this parameter for all bivariate subspaces of our system using numerical trajectories generated by varying $\theta$ (fixed $\tau_e$) and $\epsilon$. The results are shown in Fig.~\ref{fig:TE_asy}. We find that the asymmetry parameter strongly depends on these two parameters and the choice of the subspace. 
Interestingly, the asymmetry gets slightly lowered 
for $(\lambda, x_1)$ space as $\epsilon$ is enhanced for a fixed $\theta$ (Fig.~\ref{fig:TE_asy}(a)). On the contrary, this dependency is reversed in both
$(x_1, x_2)$ space (Fig.~\ref{fig:TE_asy}(b)),  
and $(\lambda, x_2)$ space (Fig.~\ref{fig:TE_asy}(c)) -  where the asymmetry is higher with higher $\epsilon$ for any fixed $\theta$. Such features of asymmetry suggest that while higher $\epsilon$ induces nonequilibrium character in $(x_1, x_2)$ and   $(\lambda, x_2)$ spaces, it reduces the same in  $(\lambda, x_1)$ space. This is consistent with our observation regarding the dependence of pairwise mutual information between different variables and further substantiates the counterintuitive role of hydrodynamic interaction in shaping the irreversibility of the process at different subspaces.  Moreover, the theoretically estimated partial entropy production rates at these (coarse-grained) subspaces also exhibit similar signatures as a function of $\theta$ and $\epsilon$, as shown in Fig.~\ref{fig:reduced_epr} of \textit{Appendix}~\ref{ap: coarse_grained_epr}.

Note that, in contrast, at the equilibrium limit (in the absence of external driving), both the time-delayed informational measures possess a symmetric feature (shown in Fig.~\ref{fig:equilibrium} of \textit{Appendix}~\ref{ap:info_quant_equilibrium}) such that $MI_{x_1\rightarrow x_2}(\tau)  \approx MI_{x_2\rightarrow x_1}(\tau) $  and  $TE_{x_1\rightarrow x_2}(\tau)  \approx TE_{x_2\rightarrow x_1}(\tau) $, which indicates a strong bidirectional (\tikz[baseline=(x1.base), every node/.style={font=\footnotesize}, >=Stealth] {
  \node (x1) {$x_1$};
  \node (x2) [right=of x1] {$x_2$};

  \draw[->,line width=1pt, bend left=10] (x1) to (x2); 
  \draw[->, line width=1pt,bend left=10] (x2) to (x1); 
}) nature of information flow between the particles. Indeed, this is consistent with the fact that dynamic correlations in equilibrium configurations are usually symmetric i.e. $\langle x(t)y(t+\tau)\rangle = \langle y(t)x(t+\tau)\rangle$~\cite{meiners1999direct}. 

\section{conclusions}
\label{sec: conclusion}
In conclusion, we utilize information-theoretic techniques to explore the role of hydrodynamic interactions in establishing mutual dependencies between various degrees of freedom in the presence of an \textit{active} drive. Our analysis reveals that while such interactions promote informational exchange between the coupled particles, the mutual information between the external drive and the driven particle, gets slightly reduced. This explains a previously observed counterintuitive dependence of the irreversibility of this system on the strength of interactions \cite{das2024irreversibility}, especially under coarse-graining. Moreover, the emergent causal structure in every phase plane of the system -- estimated through the time-delayed information-theoretic measures -- quantifies the crucial role of the nonequilibrium drive in manifesting directional influence between the particles. These findings further underscore the effectiveness of information-theoretic measures in uncovering the nonequilibrium origin in a complex system or their ability to distinguish nonequilibrium processes from their equilibrium counterparts. 
While our study focuses on a particular type of pairwise interactions, the analysis we presented here can be further generalised to nonequilibrium systems with multiple interacting degrees of freedom, under different kinds~\cite{khadka2018active,loos2020irreversibility,poggioli2023odd,pahi2025simultaneous} and higher orders of interaction ~\cite{lynn2022decomposing,rose2024role,nicoletti2024informationprx}.    
 
Such detailed knowledge about the role of interaction could be especially informative for manipulating the kinetics of mesoscopic systems subjected to nonequilibrium environments with unprecedented control and precision \cite{suchitran2024tuning, chennakesavalu2024adaptive}. Reciprocally, analyses of the flow of information at various levels can also be utilized to elicit detailed information about the kinetics of molecular machines that function in highly nonequilibrium environments~\cite{leighton2024flow}. Another relevant problem of interest could be the control of microsystems comprising multiple particles with targeted features in a fluidic environment -- where hydrodynamic interaction is prevalent~\cite{salambo2020engineered,kim2025direct,oikawa2025experimentally, das2023enhanced}.  Additionally, the prospect of achieving dynamical information synergy~\cite{hahn2023dynamical} in these systems through hydrodynamic interactions by harnessing such information flow structures can also be explored. We plan to investigate some of these aspects in future work.

\section*{Acknowledgements}
The work is supported by IISER Kolkata and the Science and Engineering Research Board (SERB), Department of Science and Technology, Government of India, through the research grant CRG/2022/002417. BD is thankful to the Ministry Of Education of
Government of India for financial support through the Prime Minister’s Research Fellowship (PMRF) grant. SKM acknowledges the Knut and Alice Wallenberg Foundation for financial support through Grant No. KAW 2021.0328.


%

\begin{widetext}
\begin{appendix}
\renewcommand{\thefigure}{A\arabic{figure}}
\setcounter{figure}{0} 
\section{Covariance matrix}
\label{ap:cov_matrix}
The elements of the covariance matrix ($\mathbf{C}$) of the steady-state distribution of the whole system can be estimated as follows,
\begin{align}
    \begin{split}
        c_{x_1x_1} &= \frac{D_0 \tau _1 \left(-\tau _2^2 \left(\alpha ^2 \theta +\alpha +1\right)+\alpha  \tau _1^2 \left(\alpha  \left(\epsilon ^2-1\right) (\alpha  \theta +1)-1\right)+\tau _2 \tau _1 \left(\alpha  \left(\alpha  \left(\epsilon ^2-1\right) (\alpha  \theta +\theta +1)-2\right)-1\right)\right)}{\left(\tau _1+\tau _2\right) \left(\alpha  \tau _1 \left(\alpha  \left(\epsilon ^2-1\right)-1\right)-(\alpha +1) \tau _2\right)}\\
        c_{x_1x_2} &= c_{x_2x_1}= -\frac{\alpha ^2 D_0 \theta  \tau _1 \tau _2^2 \epsilon }{\left(\tau _1+\tau _2\right) \left(\alpha  \tau _1 \left(\alpha  \left(\epsilon ^2-1\right)-1\right)-(\alpha +1) \tau _2\right)}\\
         c_{x_1\lambda} &= c_{\lambda x_1} = \frac{\alpha ^2 D_0 \theta  \tau _1 \left(\alpha  \tau _1 \left(\epsilon ^2-1\right)-\tau _2\right)}{\alpha  \tau _1 \left(\alpha  \left(\epsilon ^2-1\right)-1\right)-(\alpha +1) \tau _2}\\
         c_{x_2 x_2} &= -\frac{D_0 \tau _2 \left((\alpha +1) \tau _2^2+\tau _2 \tau _1 \left(\alpha  \left(\alpha  (\theta -1) \epsilon ^2+\alpha +2\right)+1\right)+\alpha  \tau _1^2 \left(-\alpha  \epsilon ^2+\alpha +1\right)\right)}{\left(\tau _1+\tau _2\right) \left(\alpha  \tau _1 \left(\alpha  \left(\epsilon ^2-1\right)-1\right)-(\alpha +1) \tau _2\right)}\\
         c_{x_2 \lambda} &= c_{\lambda x_2} = -\frac{\alpha ^2 D_0 \theta  \tau _1 \tau _2 \epsilon }{\alpha  \tau _1 \left(\alpha  \left(\epsilon ^2-1\right)-1\right)-(\alpha +1) \tau _2}\\
         c_{\lambda \lambda} &= \alpha  D_0 \theta  \tau _1
    \end{split}
    \label{eq:ap_cov_elements}
\end{align}

\section{Mutual information}
\label{ap:MI}
The pairwise mutual information between the external drive and the driven particle  can be estimated using the corresponding submatrices of the steady-state covariance matrix (according to the Eq.~\eqref{eq: MIlambdax1}) as,
\setlength{\displaywidth}{0.8\textwidth}
\begin{align}
      MI_{\lambda x_1} = \frac{1}{2} \log_2 \left(\frac{\mathcal{P}}{\mathcal{Q}}\right),
    \end{align}   
    where,
\begin{align}
    \begin{split}
        \mathcal{P} &= \left(\alpha \tau_1 \left(\alpha \left(\epsilon^2 - 1\right) - 1\right) 
        - (\alpha + 1) \tau_2\right) 
        \left(-\tau_2^2 \left(\alpha^2 \theta + \alpha + 1\right) 
        + \alpha \tau_1^2 \left(\alpha \left(\epsilon^2 - 1\right) (\alpha \theta + 1) - 1\right)\right. \\
        &\quad + \tau_2 \tau_1 \left(\alpha \left(\alpha \left(\epsilon^2 - 1\right) (\alpha \theta + \theta + 1) - 2\right) - 1\right)\Big), \\
        \mathcal{Q} &= \alpha^2 \tau_1^3 \left(\alpha \left(\epsilon^2 - 1\right) \left(\alpha \left(-\theta + \epsilon^2 - 1\right) - 2\right) + 1\right) \\
        &\quad + \tau_2^3 (\alpha (\alpha \theta + \alpha + 2) + 1) 
        + \alpha \tau_2 \tau_1^2 \left(\alpha \left(\alpha \left(\epsilon^2 - 1\right) \left(\alpha (\theta + 1) \left(\epsilon^2 - 1\right) 
        - 2 (\theta + 2)\right)\right.\right. \\
        &\quad\left.\left.- 2 \epsilon^2 + 5\right) + 2\right) 
        + \tau_2^2 \tau_1 \left(\alpha \left(\alpha \left(-2 \alpha (\theta + 1) \left(\epsilon^2 - 1\right) 
        + \theta - \left((\theta + 2) \epsilon^2\right) + 5\right) + 4\right) + 1\right).
    \end{split}
\end{align}
Interestingly, $MI_{\lambda x_1}$ in absence of the second particle reduces to,
\begin{align}
    MI_{\lambda x_1}|_{\epsilon \rightarrow0} = \frac{1}{2} \log_2\left(\frac{(\alpha +1) \left(\alpha ^2 \theta +\alpha +1\right)}{\alpha  (\alpha  \theta +\alpha +2)+1}\right),
\end{align}
which is evidently larger than $MI_{\lambda x_1}$. 

Similarly, the mutual information between the particles can be computed (according to the Eq.~\eqref{eq: MIx1x2}) as,
\begin{align}
    MI_{x_1x_2} = \frac{1}{2}\log_2\left(\frac{\mathcal{R}}{\mathcal{S}}\right),
\end{align}
where, 

\begin{align}
    \begin{split}
    \mathcal{R} &= \left(-(\alpha +1) \tau _2^2-\tau _2 \tau _1 \left(\alpha  \left(\alpha  (\theta -1) \epsilon ^2+\alpha +2\right)+1\right)
    +\alpha  \tau _1^2 \left(\alpha  \left(\epsilon ^2-1\right)-1\right)\right)\\
    &\quad\times\left(-\tau _2^2 \left(\alpha ^2 \theta +\alpha +1\right) +\alpha  \tau _1^2 \left(\alpha  \left(\epsilon ^2-1\right) (\alpha  \theta +1)-1\right)+\tau _2 \tau _1 \left(\alpha  \left(\alpha  \left(\epsilon ^2-1\right) (\alpha  \theta +\theta +1)-2\right)-1\right)\right), \\
    \mathcal{S} &= \alpha ^2 \tau _1^4 \left(\alpha  \left(\epsilon ^2-1\right)-1\right) \left(\alpha  \left(\epsilon ^2-1\right) (\alpha  \theta +1)-1\right)\\
    &+\alpha  \tau _1^3 \tau _2 \left(\alpha  \left(\alpha  \left(-\left(\alpha ^2 \theta  \left(\epsilon ^2-1\right) \left((\theta -2) \epsilon ^2+2\right)\right)+2 \alpha  \left(\epsilon ^2-1\right) \left(-2 \theta +\epsilon ^2-1\right)+2 (\theta +3)-(\theta +6) \epsilon ^2\right)-2 \epsilon ^2+6\right)+2\right)\\
   +& (\alpha  \left(\alpha  \left(\alpha  \left(-\left(\alpha ^2 \theta  \left(\epsilon ^2-1\right) \left((\theta -1) \epsilon ^2+1\right)\right)-\alpha  \left(\epsilon ^2-1\right) \left(5 \theta +\left(\theta ^2-1\right) \epsilon ^2+1\right)+5 \theta -2 (\theta +3) \epsilon ^2+6\right)+\theta -4 \epsilon ^2+10\right)+6\right)+1) \\
   &\times \tau _1^2 \tau _2^2 + (\alpha +1) \tau _2^4 \left(\alpha ^2 \theta +\alpha +1\right)+\tau _1 \tau _2^3 \left(\alpha  \left(\alpha  \left(\alpha  \left(2 (\alpha +2) \theta -\left(\epsilon ^2 (2 \alpha  \theta +\theta +2)\right)+2\right)+2 \theta -2 \epsilon ^2+6\right)+6\right)+2\right).
    \end{split}
\end{align}
Note that, if the particles are well separated such that $\epsilon\rightarrow0$, $MI_{x_1x_2}$ also vanishes. 
Furthermore, the dependence between the fixed particle ($x_2$) and the external drive ($\lambda$) can also be estimated through their mutual information as,
\begin{align}
    MI_{\lambda x_2} = \frac{1}{2}\log_2\frac{c_{\lambda \lambda}c_{x_2x_2} }{\det \mathbf{C}_{\lambda x_2}}=\frac{1}{2}\log_2\left(\frac{\mathcal{T}}{\mathcal{U}}\right),
\end{align}
where,
\begin{align}
    \begin{split}
        \mathcal{T} &= \left(\alpha  \tau _1 \left(\alpha  \left(\epsilon ^2-1\right)-1\right)-(\alpha +1) \tau _2\right) \left(-(\alpha +1) \tau _2^2-\tau _2 \tau _1 \left(\alpha  \left(\alpha  (\theta -1) \epsilon ^2+\alpha +2\right)+1\right)+\alpha  \tau _1^2 \left(\alpha  \left(\epsilon ^2-1\right)-1\right)\right), \\
        \mathcal{U} &= \alpha ^2 \tau _1^3 \left(-\alpha  \epsilon ^2+\alpha +1\right)^2+(\alpha +1)^2 \tau _2^3+\alpha  \tau _2 \tau _1^2 \left(\alpha  \left(-\alpha  \left(\epsilon ^2-1\right) \left(\alpha  (\theta -1) \epsilon ^2+\alpha +4\right)-2 \epsilon ^2+5\right)+2\right)\\
        &\quad +\tau _2^2 \tau _1 \left(\alpha  \left(\alpha  \left(-2 \alpha  \left(\epsilon ^2-1\right)+(\theta -2) \epsilon ^2+5\right)+4\right)+1\right). 
    \end{split}
\end{align}
Importantly, $MI_{\lambda x_2} \neq 0$ as long as $\epsilon \neq0$ and $MI_{\lambda x_2} < MI_{\lambda x_1}$, which indicates that the external driving is partially available to the fixed particle via hydrodynamic interaction.

\section{Details of numerical simulation of the experimental system}
\label{ap:numerical_Model}
The coupled \textit{Langevin} equations that describe the dynamics of our system are provided in  Eq.~\eqref{eq:ap_dyn_neq1}. Noticeably, the diffusion matrix ($\mathbf{D}$) for this system is not diagonal due to its connection with the hydrodynamic coupling tensor, so the matrix incorporating the strength of the appropriate noise terms ($\mathbf{G}$) is calculated after performing Cholesky decomposition such that $\mathbf{D} = \frac{1}{2} \mathbf{GG}^T$~\cite{das2024irreversibility}.   
The noise matrix $\mathbf{G}$ of our system takes the following form, 
\begin{equation}
   \mathbf{G} = \begin{pmatrix}
    \sqrt{2 D_0}& 0& 0\\
    \epsilon \sqrt{2 D_0} & \sqrt{2(D_0-\epsilon^2D_0)} & 0\\
    0& 0& \sqrt{2 \theta D_0}   
\end{pmatrix}. 
\end{equation}

Then the linear, stochastic differential equations with the drift  (Eq.~\eqref{eq:drift_diffusive}) and noise matrices are numerically discretized with fixed time step $\Delta t = 1\times10^{-4}s$ (which is less than all timescales present in the systems) and the 1st-order Eurler-Maruyama method is used to perform the integration. 
The discretized form of the dynamical equations will be,
\begin{align}
    \begin{split}
        x_1^{t+\Delta t} &= x_1^t - (1/\tau_1)\ x_1^t \Delta t - (\epsilon /\tau_2)\ x_2^t \Delta t + (1/\tau_1)\ \lambda^t \Delta t + \sqrt{2D_0 \Delta t}\ \eta_1^t, \\
        x_2^{t+\Delta t} &= x_2^t - (\epsilon /\tau_1)\ x_1^t\Delta t - (1/\tau_2)\ x_2^t\Delta t + (\epsilon/\tau_1)\ \lambda^t\Delta t  + \epsilon \sqrt{2D_0 \Delta t}\ \eta_1^t + \sqrt{2(D_0-\epsilon^2D_0)\Delta t }\ \eta_2^t, \\
        \lambda^{t + \Delta t} &= \lambda^t -(1/(\alpha \tau_1))\ \lambda^t \Delta t +  \sqrt{2 \theta D_0 \Delta t}\ \eta_3^t,
    \end{split}
    \label{ap:eq_ou_lan_discrete}
\end{align}
where $\eta_i^t \sim \mathcal{N}(0,1)$ is sampled from a Gaussian distribution with zero mean and unit variance and the initial states ($x_1^0, x_2^0, \lambda^0$) of the individual degrees of freedom are sampled from normal distributions with zero mean and appropriate variances such that, $x_1^0\sim \mathcal{N}(0,\sigma_1^2)$,   $x_2^0\sim \mathcal{N}(0,\sigma_2^2)$ and $\lambda^0\sim \mathcal{N}(0,\sigma_\lambda^2)$ with $\sigma_{1,2} = \sqrt{k_B T/k_{1,2}}$ and $\sigma_{\lambda} = \sqrt{D_e/\tau_e}$. The system parameters ($\tau_1$, $\tau_2$, $\tau_e$, $D_0$ and $\alpha$) remain the same as mentioned in the \textit{Experimental Model} section (Section ~\ref{sec:theory_model}) for every numerical simulation.   \\

\section{Time-delayed informational measures for other \texorpdfstring{$\theta$}{theta} and \texorpdfstring{$\epsilon$}{epsilon} }
\label{ap:TE_MI_both}
\begin{figure*}
    \centering
    \includegraphics[width=0.9\linewidth]{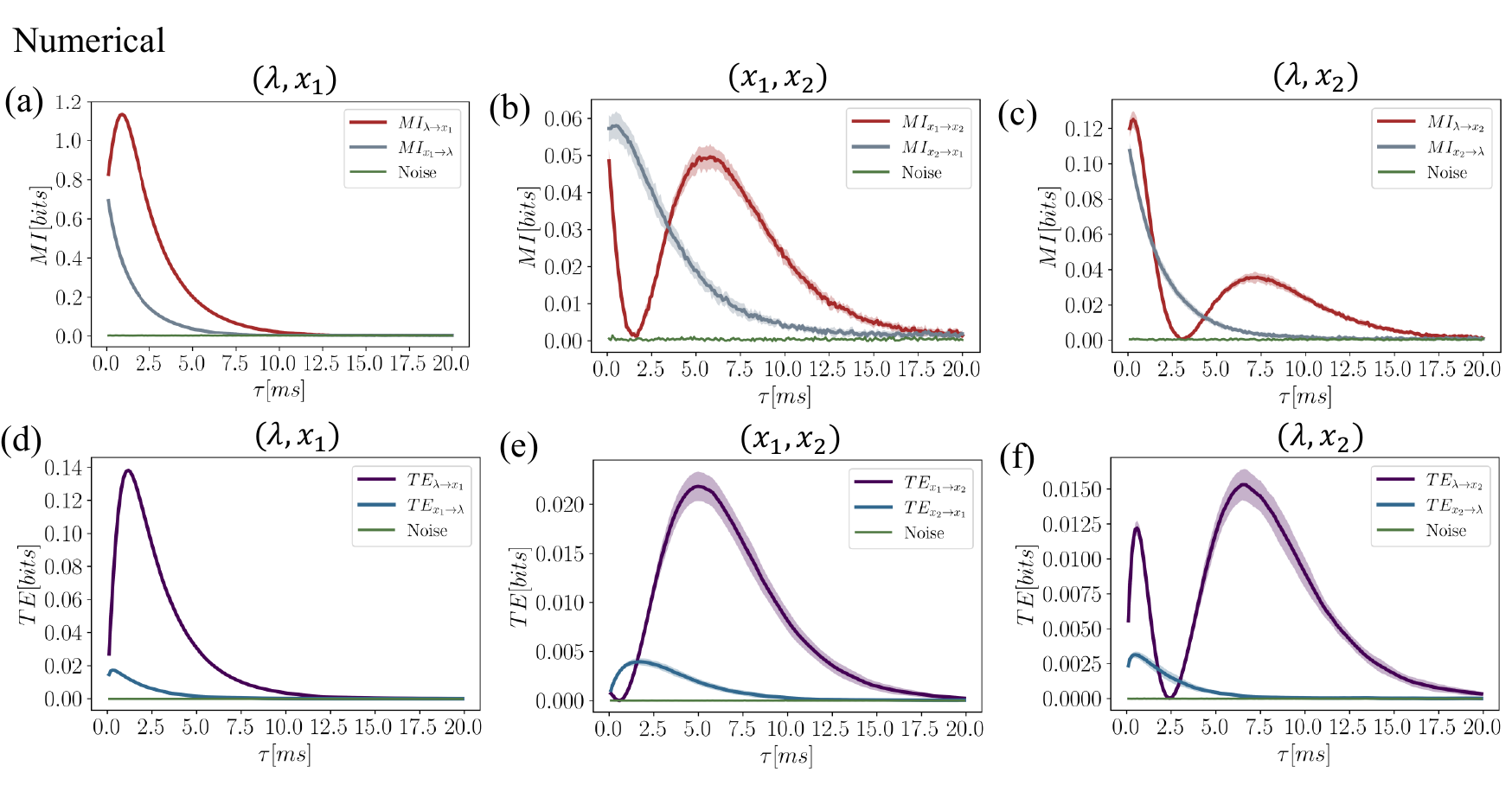}
    \caption{Estimation of time-delayed informational measures for the system with $\theta = 5.6$ and $\epsilon = 0.49$: (a)-(c) time-delayed mutual informations, (d)-(f) transfer entropies between different variables are estimated in (a) \& (d) $(\lambda,x_1)$, (b) \& (e) $(x_1,x_2)$ and (c) \& (f) $(\lambda,x_2)$ spaces.}
    \phantomsection
    \label{fig:MI_TE_highTheta}
\end{figure*}
We estimate the time-delayed mutual information and the transfer entropy for the system with higher nonequilibrium strength $\theta = 5.6$ from numerical trajectories. As we show in Fig.~\ref{fig:MI_TE_highTheta}, the unidirectionality of information flow in the hierarchical dynamics for higher activity strength can be observed from the pronounced asymmetry of transfer entropy estimated between different variables of the system as $TE_{\lambda \rightarrow x_1} \gg TE_{x_1 \rightarrow \lambda}$, $TE_{x_1 \rightarrow x_2} \gg TE_{x_2 \rightarrow x_1}$ and $TE_{\lambda \rightarrow x_2} \gg TE_{x_2 \rightarrow \lambda}$.  Additionally, the magnitude of time-delayed measures also depends on the interaction strength $\epsilon$ while the strength of external drive $\theta$ is fixed. As displayed in Fig.~\ref{fig:MI_TE_loweps}, the peak magnitude of $TE_{\lambda \rightarrow x_1}$ is slightly higher than the same for $\epsilon = 0.49$ (Fig.~\ref{fig:TE_time}), while the peak magnitudes of  $TE_{x_1 \rightarrow x_2}$ and $TE_{\lambda \rightarrow x_2}$ gets reduced comparing the values for $\epsilon = 0.49$ (Fig.~\ref{fig:TE_time}). The same can be further understood from the inhomogeneous features of asymmetry parameters as a function of $\theta$ and $\epsilon$ in every bivariate subspace of our system, as shown in Fig.~\ref{fig:TE_asy}.   
\begin{figure*}
    \centering
    \includegraphics[width=0.9\linewidth]{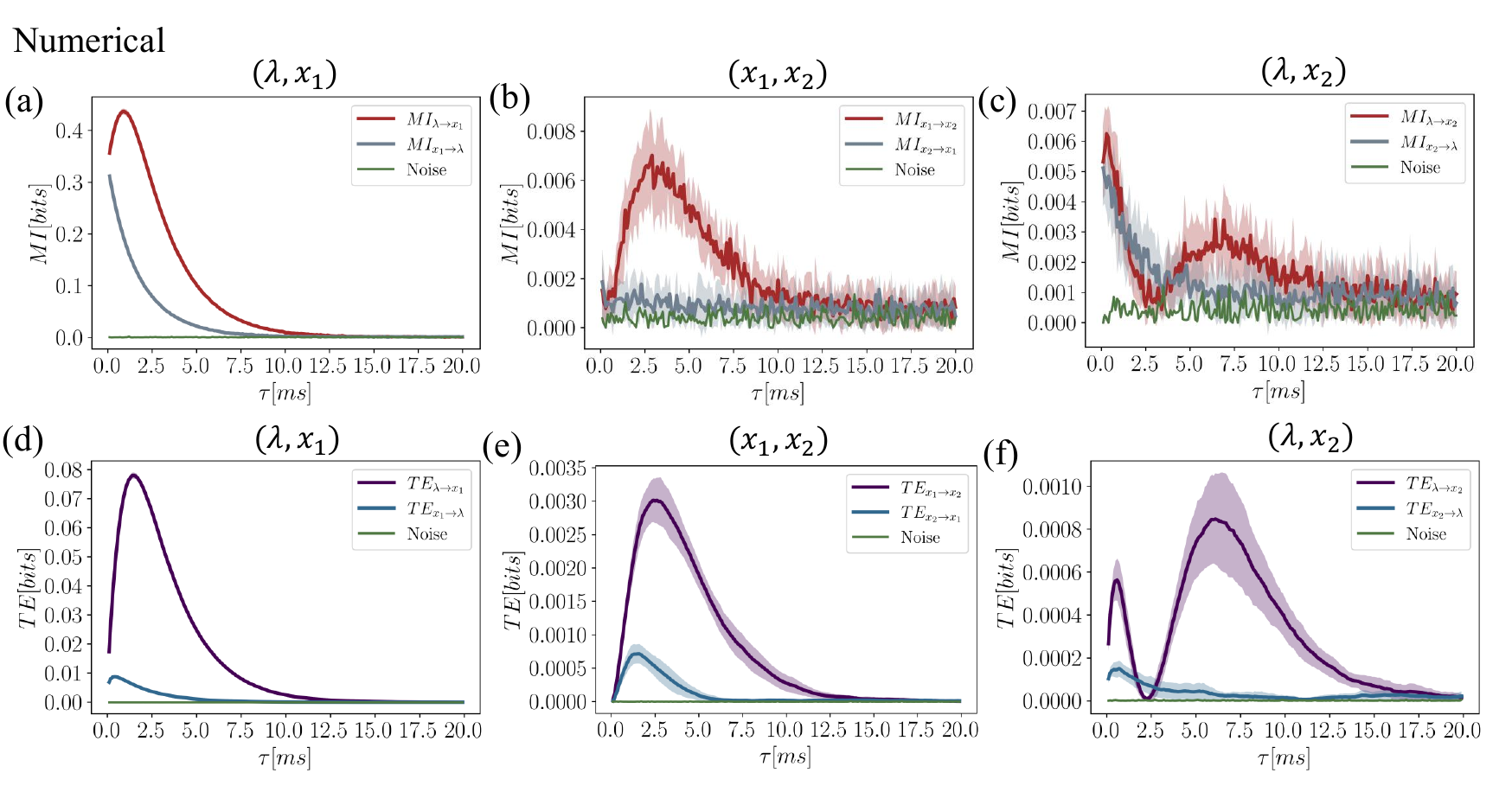}
    \caption{Estimation of time-delayed informational measures for the system with ($\theta = 0.56$, $\epsilon = 0.29$): (a)-(c) time-delayed mutual information, (d)-(f) transfer entropy between different variables are estimated in (a) \& (d) $(\lambda,x_1)$, (b) \& (e) $(x_1,x_2)$ and (c) \& (f) $(\lambda,x_2)$ spaces.}
    \phantomsection
    \label{fig:MI_TE_loweps}
\end{figure*}

\section{Time-dependent cross-correlation functions}
\label{ap:ccf}
The two-time correlation matrix $\langle \mathbf{x}(t)\mathbf{x}(t+\tau)\rangle$ corresponding to the system as a function of time lag $\tau (> 0)$ can be calculated as~\cite{villamaina2009fluctuation,das2023enhanced},
\begin{equation}
\label{ap:eq_ccf}
    \mathbf{C}_{\mathbf{xx}}(\tau) \equiv \langle \mathbf{x}(t)\mathbf{x}(t+\tau)\rangle = [\exp[{- \mathbf{F} \tau]\mathbf{C}}]^T.
\end{equation}
Furthermore, each two-time cross-correlation function can be normalised, e.g. $ {NC}_{x_{1}x_{2}}(\tau) = \frac{\langle x_1(t) x_2(t+\tau)\rangle}{\sqrt{\text{Var}(x_1)} \sqrt{\text{Var}(x_2)}} \equiv \frac{\langle x_1(t) x_2(t+\tau)\rangle}{\sqrt{c_{x_{1}x_{1}}} \sqrt{c_{x_{2}x_{2}}}}$, $ {NC}_{\lambda x_{1}}(\tau) = \frac{\langle \lambda(t) x_1(t+\tau)\rangle}{\sqrt{\text{Var}(\lambda)} \sqrt{\text{Var}(x_1)}} \equiv \frac{\langle \lambda(t) x_1(t+\tau)\rangle}{\sqrt{c_{\lambda \lambda}} \sqrt{c_{x_{1}x_{1}}}}$, $ {NC}_{\lambda x_{2}} (\tau)= \frac{\langle \lambda(t) x_2(t+\tau)\rangle}{\sqrt{\text{Var}(\lambda)} \sqrt{\text{Var}(x_2)}} \equiv \frac{\langle \lambda(t) x_2(t+\tau)\rangle}{\sqrt{c_{\lambda \lambda}} \sqrt{c_{x_{2}x_{2}}}}$ 
 and so on.\\ 

We do not provide detailed analytical expressions for each correlation function for brevity. However, we plot the normalised cross-correlation functions(NCCF) corresponding to the dynamical variables in Fig.~\ref{fig:nccf},  for a particular set of fixed parameters: $D_0 \sim 0.16~\mu m^2/s$, $\tau_1 \sim 1.5~ms$, $\tau_2 \sim 2.1~ms$,  $\alpha\sim2.7$. It is quite evident that the two-time correlation functions corresponding to any two dynamical variables are asymmetric in the presence of nonequilibrium drive as $NC_{\lambda x_{1}}(\tau) \neq NC_{x_{1}\lambda}(\tau)$, $NC_{x_{1}x_{2}}(\tau) \neq NC_{x_{2}x_{1}}(\tau)$ and  $NC_{\lambda x_{2}}(\tau) \neq NC_{x_{2}\lambda}(\tau)$. These asymmetric features are enhanced with the increase in nonequilibrium strength~\cite{ohga2023thermodynamic}. In this context, the dynamics of the driven particle ($x_1(t)$) are positively correlated with the external nonequilibrium drive ($\lambda(t)$) and $NC_{\lambda x_{1}}(\tau)$ possesses a global maxima at a particular delay time (Fig.~\ref{fig:nccf}(a) and \ref{fig:nccf}(d)). On the other hand, the trapped particles are anti-correlated and the corresponding (anti-) correlations ($NC_{x_{1} x_{2}}(\tau)$ and $NC_{x_{2} x_{1}}(\tau)$) also peaks at a particular time lag (Fig.~\ref{fig:nccf}(b) and \ref{fig:nccf}(e)). Interestingly, as discussed in the main text, the corresponding time-delayed information-theoretic quantities (time-delayed mutual information and transfer entropy) also display global maxima at these particular time lags. Moreover, the correlation functions between the external drive and the fixed particle, $NC^\tau_{\lambda x_2}$ shows both maximum positive correlation and maximum negative correlation at two different time lags (Fig.~\ref{fig:nccf}(c) and \ref{fig:nccf}(f)). As a result, both time-delayed mutual information ($MI_{\lambda \rightarrow x_2}(\tau) $) and transfer entropy ($TE_{\lambda \rightarrow x_2}(\tau) $) between these two variables also show two peaks at two different time lags.        
\begin{figure*}[t]
    \centering
    \includegraphics[width=0.9\linewidth]{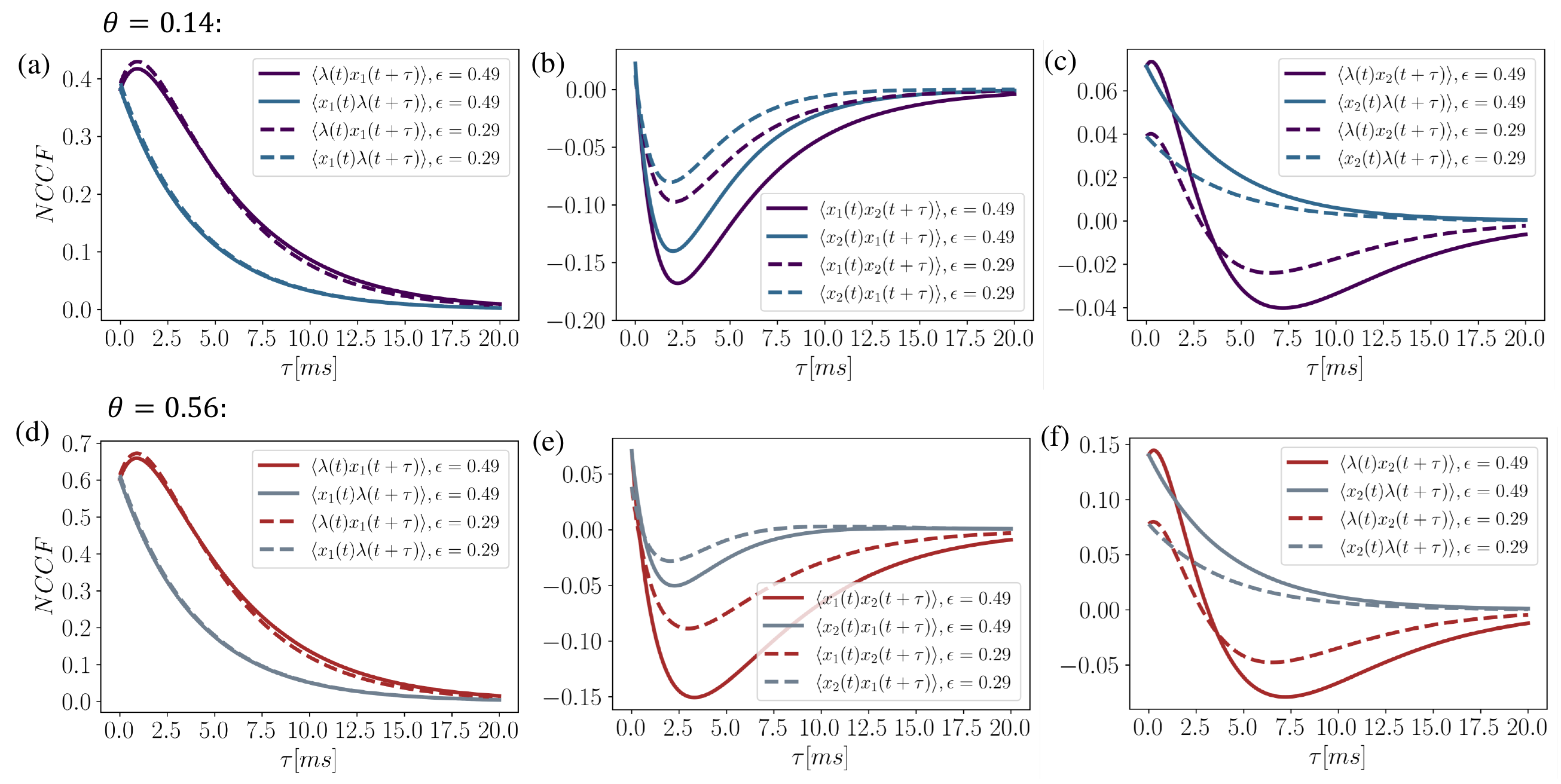}
    \caption{Analytically estimated normalised cross-correlation function (NCCF) corresponding to the dynamical variables for two different noise strengths ($\theta$). (a)-(c): $\theta = 0.14$, (d)-(f): $\theta = 0.56$.}
    \phantomsection
    \label{fig:nccf}
\end{figure*}

\section{Analytical estimation of time-delayed mutual information}
\label{ap:tdmi_analytics}
For a stochastic system with a multivariate Gaussian distribution, the time-delayed mutual information between different degrees of freedom can be estimated analytically as well.  For any two such variables $x$ and $y$ - whose joint distribution is bivariate Gaussian, the time-delayed mutual information $MI_{x \rightarrow y}(\tau)$
defined in Eq.~\eqref{eq:MI_def_timelagged} of the main text can be analytically rewritten as,
\begin{equation}
    MI_{x \rightarrow y}(\tau) \equiv I(x_t;y_{t+\tau})= \frac{1}{2} \log_2 \frac{c_{xx} c_{yy}}{\det \mathbf{C}_{xy}(\tau)} 
\end{equation}
following its time-independent counterpart. Here, $\mathbf{C}_{xy}(\tau)$ is the time-delayed covariance matrix defined as, 
\begin{equation}
    \mathbf{C}_{xy}(\tau) = \begin{pmatrix}
        \langle x_t x_t\rangle & \langle x_t y_{t+\tau}\rangle \\
        \langle y_{t+\tau} x_t \rangle & \langle y_{t+\tau} y_{t+\tau}\rangle
    \end{pmatrix} = \begin{pmatrix}
        c_{xx} & c_{xy}(\tau) \\
        c_{xy}(\tau) & c_{yy}
    \end{pmatrix}.
\end{equation}
Hence, $MI_{x \rightarrow y}(\tau)$ can be analytically represented through the elements of $\mathbf{C}_{xy}(\tau)$ as,
\begin{equation}
    MI_{x \rightarrow y}(\tau) = \frac{1}{2} \log_2 \frac{c_{xx} c_{yy}}{c_{xx} c_{yy} - c_{xy}(\tau)^2}. 
\end{equation}
It is now clear that if $c_{xy}(\tau) \neq c_{yx}(\tau)$, $MI_{x \rightarrow y}(\tau) \neq  MI_{y \rightarrow x}(\tau)$ - suggesting a preferred directionality of information flow. Following these, we can analytically estimate $MI_{\lambda \rightarrow x_1}(\tau)$, $MI_{x_1 \rightarrow x_2}(\tau)$, $MI_{\lambda \rightarrow x_2}(\tau)$ along with their reverse counterparts using the elements of covariance matrix (as given in Eqs.~\eqref{eq:ap_cov_elements}) and the elements of two-time correlation matrix (Eq.~\eqref{ap:eq_ccf}). As shown in Fig.~\ref{fig:TDMI_analytics}, the numerically computed time-delayed mutual information reasonably matches with the corresponding analytical estimations.  This further suggests that the particular time-lags at which maxima occur for the time-delayed mutual information are indeed related to the time-lag of extremum two-point (anti-) correlation.   
\begin{figure*}
    \centering
    \includegraphics[width=0.9\linewidth]{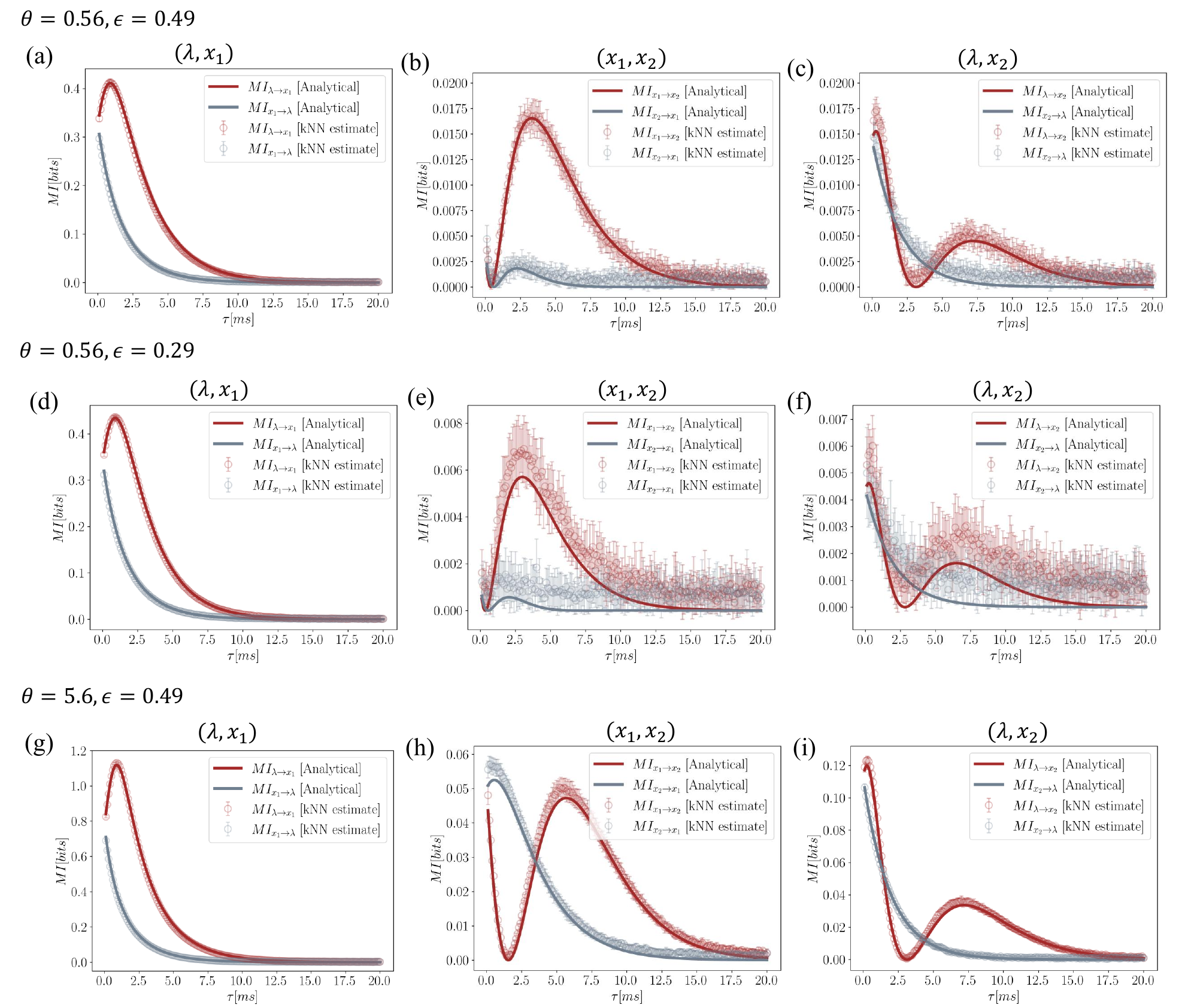}
    \caption{Analytically estimated time-delayed mutual information in different bivariate planes, along with the corresponding kNN estimates, are shown for different values of $\theta$ and $\epsilon$. }
    \phantomsection
    \label{fig:TDMI_analytics}
\end{figure*}
\section{Partial entropy production rate at bivariate planes}
\label{ap: coarse_grained_epr}
\begin{figure*}
    \centering
    \includegraphics[width=0.95\linewidth]{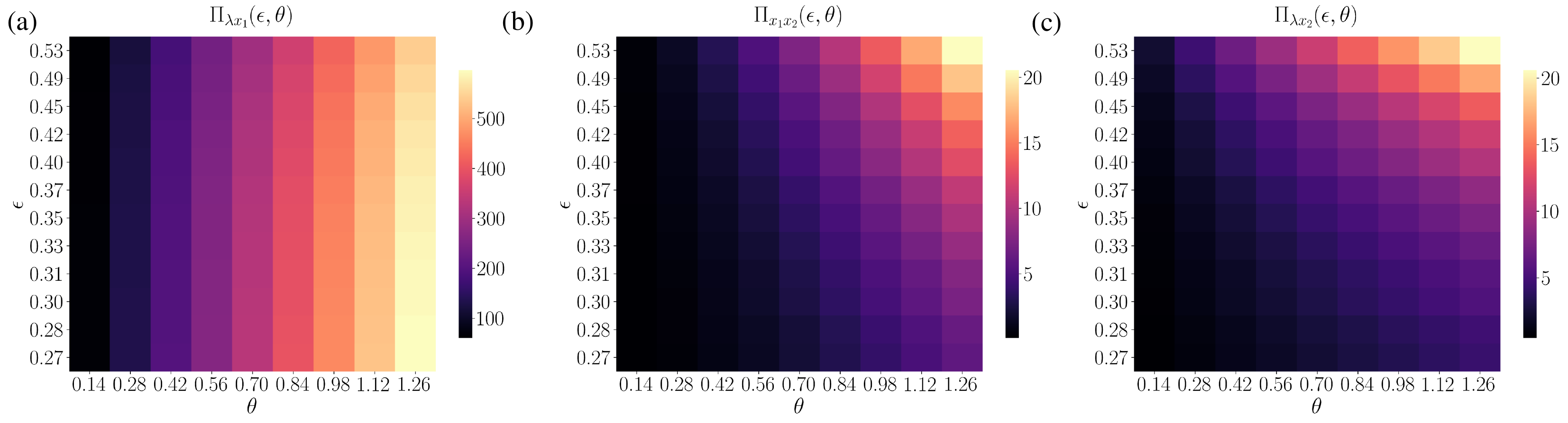}
    \caption{Theoretically estimated partial entropy production rates (in units of $k_B/s$) as a function of $\theta$ and $\epsilon$, are shown for the (a) $(\lambda, x_1)$, (b) $(x_1, x_2)$, and $(\lambda, x_2)$ spaces, respectively.  }
    \phantomsection
    \label{fig:reduced_epr}
\end{figure*}
The partial entropy production rate for each bivariate phase plane of this system can be estimated analytically using the methodology described in Ref.~\cite{nicoletti2024information}. While we demonstrated this explicitly for the $(x_1,x_2)$ phase plane in Ref.~\cite{das2024irreversibility}, the same analysis can be extended to other phase planes as well. The partial entropy production rates corresponding to all bivariate planes (coarse-grained subspaces) of our system are illustrated in Fig.~\ref{fig:reduced_epr}. Interestingly, the features of such partial or reduced entropy production rates in each plane show similar dependence on $\epsilon$ and $\theta$ as that observed in the asymmetry of transfer entropies in corresponding phase planes (Fig.~\ref{fig:TE_asy}). This further highlights the intricate connection between irreversibility and the asymmetry of transfer entropies.

\section{Comparison with the equilibrium scenario }
\label{ap:info_quant_equilibrium}
In the absence of any external drive, the hydrodynamically coupled system will reach an equilibrium steady state in which the particles are statistically independent. Since the steady-state joint distribution of the displacement of the particles is exactly factorized as the product of marginals i.e. $P(x_1,x_2) = P(x_1)P(x_2)$ (as shown in Fig.~\ref{fig:equilibrium}(a)), mutual information at equilibrium steady state will be zero. Interestingly, the time-delayed informational measures possess a symmetric feature such that $MI_{x_1\rightarrow x_2}(\tau) \approx MI_{x_2\rightarrow x_1}(\tau)$ (Fig.~\ref{fig:equilibrium}(b)) and  $TE_{x_1\rightarrow x_2}(\tau) \approx TE_{x_2\rightarrow x_1}(\tau)$ (Fig.~\ref{fig:equilibrium}(c)), which indicates a strong bidirectional (\tikz[baseline=(x1.base), every node/.style={font=\footnotesize}, >=Stealth] {
  \node (x1) {$x_1$};
  \node (x2) [right=of x1] {$x_2$};

  \draw[->,line width=1pt, bend left=10] (x1) to (x2); 
  \draw[->, line width=1pt,bend left=10] (x2) to (x1); 
}) nature of information flow for the equilibrium configuration. The characteristics of the information flow in an equilibrium scenario are quite different than the same in a nonequilibrium configuration and consistent with the fact that dynamic correlations in equilibrium configurations are usually symmetric i.e. $\langle x(t)y(t+\tau)\rangle = \langle y(t)x(t+\tau)\rangle$. 
\begin{figure*}[h]
    \centering
    \includegraphics[width=0.9\linewidth]{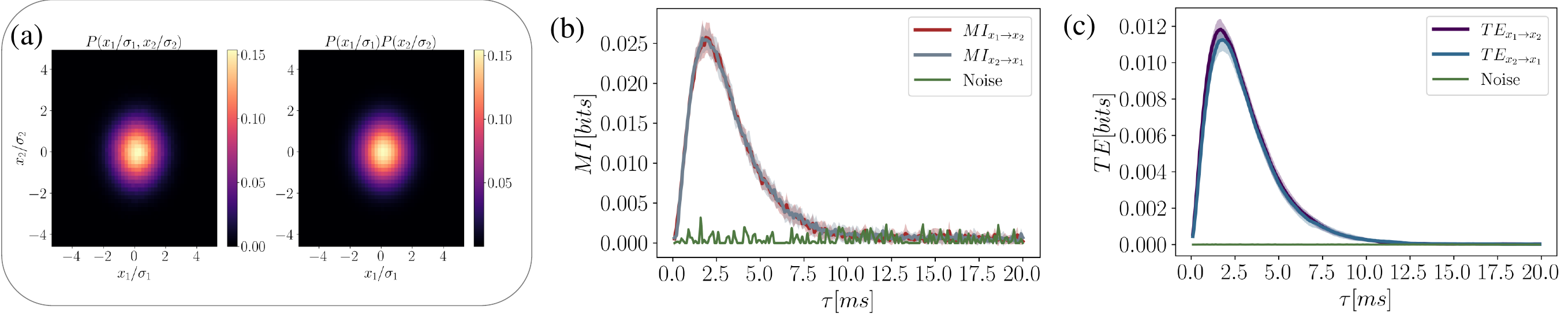}
    \caption{(a) Joint probability distribution of the displacements of the particles at equilibrium steady state can be exactly factorized as the product of the marginals. (b) Time-delayed mutual information and (c) transfer entropy between the particles ($(x_1,x_2)$ space) with hydrodynamic interaction $\epsilon = 0.49$ are estimated from the numerical trajectories of the system without a nonequilibrium drive. The `green' line in each plot denotes the estimates from the randomly shuffled trajectories.}
    \phantomsection
    \label{fig:equilibrium}
\end{figure*}

\end{appendix}
\end{widetext}
\end{document}